\documentclass[aps,PRX,twocolumn,amsmath,amssymb,floatfix]{revtex4-2}
\usepackage{CJK}
\usepackage{graphicx}
\usepackage{dcolumn}
\usepackage{bm}
\usepackage{xcolor}
\usepackage[T1]{fontenc}
\usepackage{lmodern}
\usepackage[colorlinks=true,linkcolor=blue,urlcolor=blue,citecolor=blue]{hyperref}

\begin{document}


\title{The origin of large amplitude oscillations of dust particles in a plasma sheath}

\author{Joshua M\'endez Harper}%
\email[M\'endez Harper ]{joshua.mendez@emory.edu}
\author{Guram Gogia}%
\author{Brady Wu}%
\author{Zachary Laseter}%
\author{Justin C. Burton}%
\affiliation{Department of Physics, Emory University}
\date{\today}

\begin{abstract}
Micron-size charged particles can be easily levitated in low-density plasma environments. At low pressures, suspended particles have been observed to spontaneously oscillate around an equilibrium position. In systems of many particles, these oscillations can catalyze a variety of nonequilibrium, collective behaviors. Here, we report spontaneous oscillations of single particles that remain stable for minutes with striking regularity in amplitude and frequency. The oscillation amplitude can also exceed 1 cm, nearly an order of magnitude larger than previously observed. Using an integrated experimental and numerical approach, we show how the motion of an individual particle can be used to extract the electrostatic force and equilibrium charge variation in the plasma sheath. Additionally, using a delayed-charging model, we are able to accurately capture the nonlinear dynamics of the particle motion, and estimate the particle's equilibrium charging time in the plasma environment.

\end{abstract}

\maketitle

\section{Introduction}

\begin{figure}[!]
	\centering
	\includegraphics[width=3.2 in]{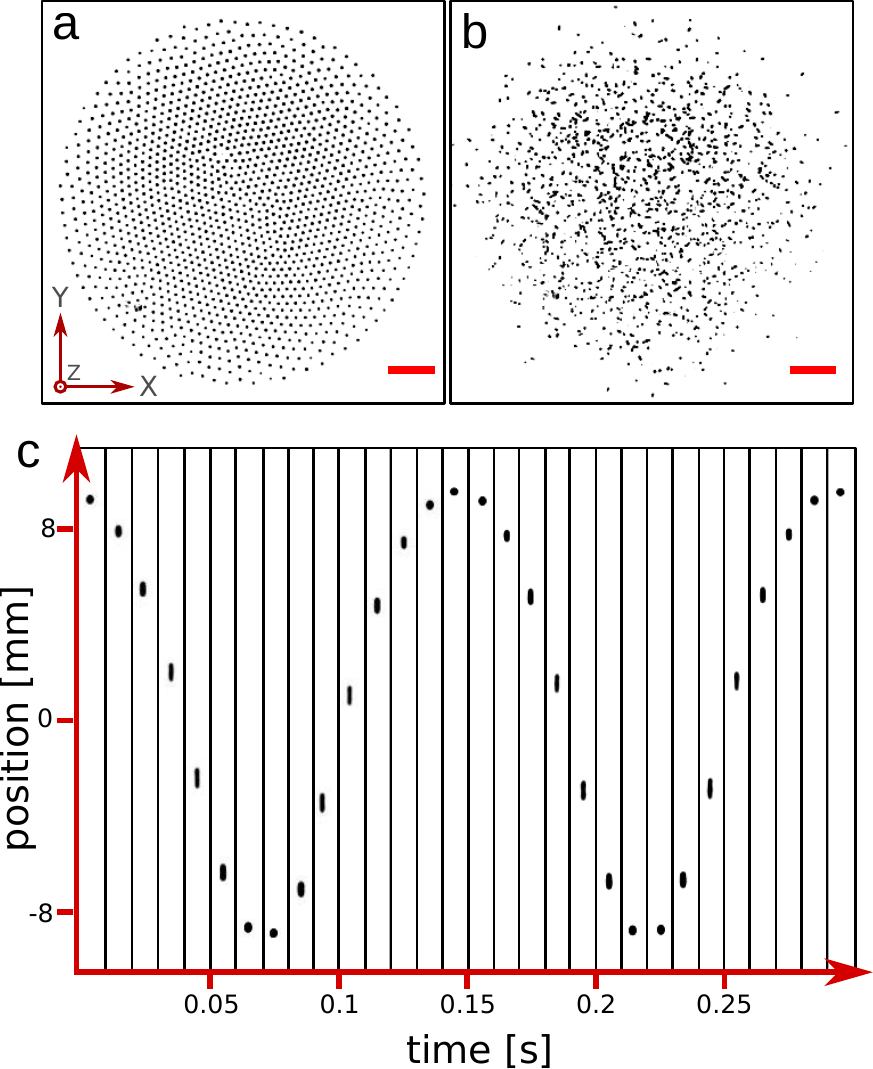}
	\caption{\label{crystal_gas} Images showing the existence of both crystalline (a) and disordered, gas-like (b) states. The scale bars are 5 mm. (c) Sequential frames from a video showing large amplitude vertical oscillations of a single particle. All images have been processed for clarity. The resolution of the camera was 23 pixels/mm. }
\end{figure}

Dusty or complex plasmas, consisting of microscopic particles immersed in a weakly-ionized gas, are ubiquitous in the natural universe and play important roles in man-made systems \cite{Boufendi2002,Merlino2006,Wahlund2009, Mendez2018KCL,Mendez2018, Mendez2019effect}. When exposed to a plasma environment, particles gain charge by collecting discrete ions and electrons on their surfaces. Because the electrons in the plasma have a substantially higher temperature than the ions do, particles attain a negative charge - usually thousands to tens of thousands of elementary charges.  The interaction energy of neighboring particles often supersedes their thermal energy, yielding a strongly coupled system that, depending on the confinement strength, can self-organize into crystalline (Fig.\ \ref{crystal_gas}a) or gas-like (Fig.\ \ref{crystal_gas}b) states \cite{Thomas1994, Chu1994}. 

Because of the weak hydrodynamic dissipation of the surrounding neutral gas, the dust particles also experience under-damped dynamics, instabilities \cite{Piel2001}, collective vibrational modes \cite{Couedel2010,Liu2010}, and manifest various nonequilibrium phenomena \cite{Gogia2017,Wong2018}. Perhaps the simplest illustration of this behavior is the spontaneous vertical oscillation of particles suspended in an rf plasma sheath. Such phenomena has been observed and described by a multitude of authors over the last 20 years  \cite{Nunomura1999,Ivlev2000Inf,Takamura2001,Samarian2001,Vaulina2001,Resendes2002,Sorasio2002,Sorasio2002b}. These oscillations seem to initiate once the gas pressure is reduced below a threshold value. The amplitude of motion, reported previously to be up to a few millimeters or less, increases with decreasing pressure. 

A number of mechanisms have been proposed and explored to explain the origin of the vertical oscillations \cite{Sorasio2002b}. Nunomura et al. \cite{Nunomura1999} presented a delayed-charging mechanism that requires the equilibrium charge on a particle, $Q_\text{eq}$, to increase with height in the sheath (i.e. $\text{d}Q_\text{eq}/\text{d}z>0$), leading to an effective ``negative damping.'' The threshold for this mechanism was given by Ivlev et al. \cite{Ivlev2000Inf}, who also showed that stochastic charge fluctuations can parametrically couple to delayed charging to induce oscillations for small particles ($\approx$ 1 $\mu$m in diameter). Samarian et al. \cite{Samarian2001} observed oscillations at higher pressures and suggested both spatial variation of charge \cite{Vaulina2001} on the dust and boundary effects near the rf electrode. Finally, Resendes et al. \cite{Resendes2002} and Sorasio et al. \cite{Sorasio2002} suggested a model based on fluctuations of the plasma sheath environment. More recently, an analysis of the motion of single particles showed a strong reduction in the effective damping rate, consistent with a delayed-charging mechanism \cite{Pustylnik2006}. Despite these laudable efforts, a lack of robust experimental data for the particle motion, coupled with the inherent complexity of low-density plasma sheaths, has limited our understanding of these nonequilibrium oscillations and their associated nonlinear dynamics.

Vertical oscillations have also been identified as the driving mechanism behind melting of dust plasma crystals through a ``mode coupling" instability \cite{Liu2010,Couedel2010,Zhdanov2009}. The instability is due to the presence of a ``virtual'' charge below each particle from each particle's associated ion wake field. However, this instability acts on a lattice of particles, and does not manifest as vertical oscillations of a single, isolated particle. Recurrent melting and re-crystallization of a dusty plasma crystal has also been recently observed and linked to self-induced vertical oscillations of single particles in the lattice \cite{Gogia2017}. The characteristic timescale of switching between the two states is $\sim$ 100 seconds, which is much larger than any single-particle timescale in the system (such as the inverse damping rate of the particles [$1/\gamma \sim$ 0.5-2.0 s] or the vertical oscillation period [$1/f_0 \sim$ 0.1 s]). This emergent switching phenomenon can serve as a model to study similar dynamics in more complex systems \cite{Gogia2019}. 

Interestingly, the vertical oscillations observed in Gogia et al. \cite{Gogia2017} have characteristics that, as far as we are aware, have not been observed previously. Undoubtedly, the most striking difference between that work and previous investigations involves the amplitude of oscillations. For instance, Nunomura et al. \cite{Nunomura1999} and Takamura and others \cite{Takamura2001}, who ascribe oscillations to a delayed-charging mechanism, report amplitudes on the order of hundreds of microns. In addition, Nunomura et al. \cite{Nunomura1999} showed that the amplitude of oscillation in their experiments varied considerably over a few seconds. Similarly, Samarian et al. describe oscillations that, for a wide range of pressures and plasma powers, have amplitudes no larger than 1 or 2 mm \cite{Samarian2001}. Other authors have reported comparable oscillation amplitudes \cite{Resendes2002, Sorasio2002}. Conversely, the observations described in Gogia et al. \cite{Gogia2017}) show that single, levitated micron-sized particles can oscillate with amplitudes larger than 3 mm with surprising regularity. 

In this article, we investigate the nature of such large-amplitude and regular oscillations using both experiments and numerical simulations. The observed amplitude in our experiments can exceed 1 cm, comparable to the extent of the plasma sheath, and the motion is highly anharmonic. The oscillations are initiated below a threshold pressure, typically 1 Pa, and depend on particle size and local plasma properties. We characterize the plasma environment using a Langmuir probe and find no appreciable fluctuations at low frequencies that may affect the particle motion. Our numerical simulations suggest that stochastic fluctuations are insufficient to give rise to such large amplitudes of motion. We analyze hundreds of cycles of the particles' motion in order to extract the electrostatic force and spatial dependence of the equilibrium charge. We find that the model of delayed charging presented by Ivlev et al. \cite{Ivlev2000Inf} can accurately reproduce the motion. Our results also provide a quantitative estimate of the particle charging rate, $\nu$, which is notoriously difficult to measure by other experimental methods.

\begin{figure}[!]
	\centering
	\includegraphics[width=\columnwidth]{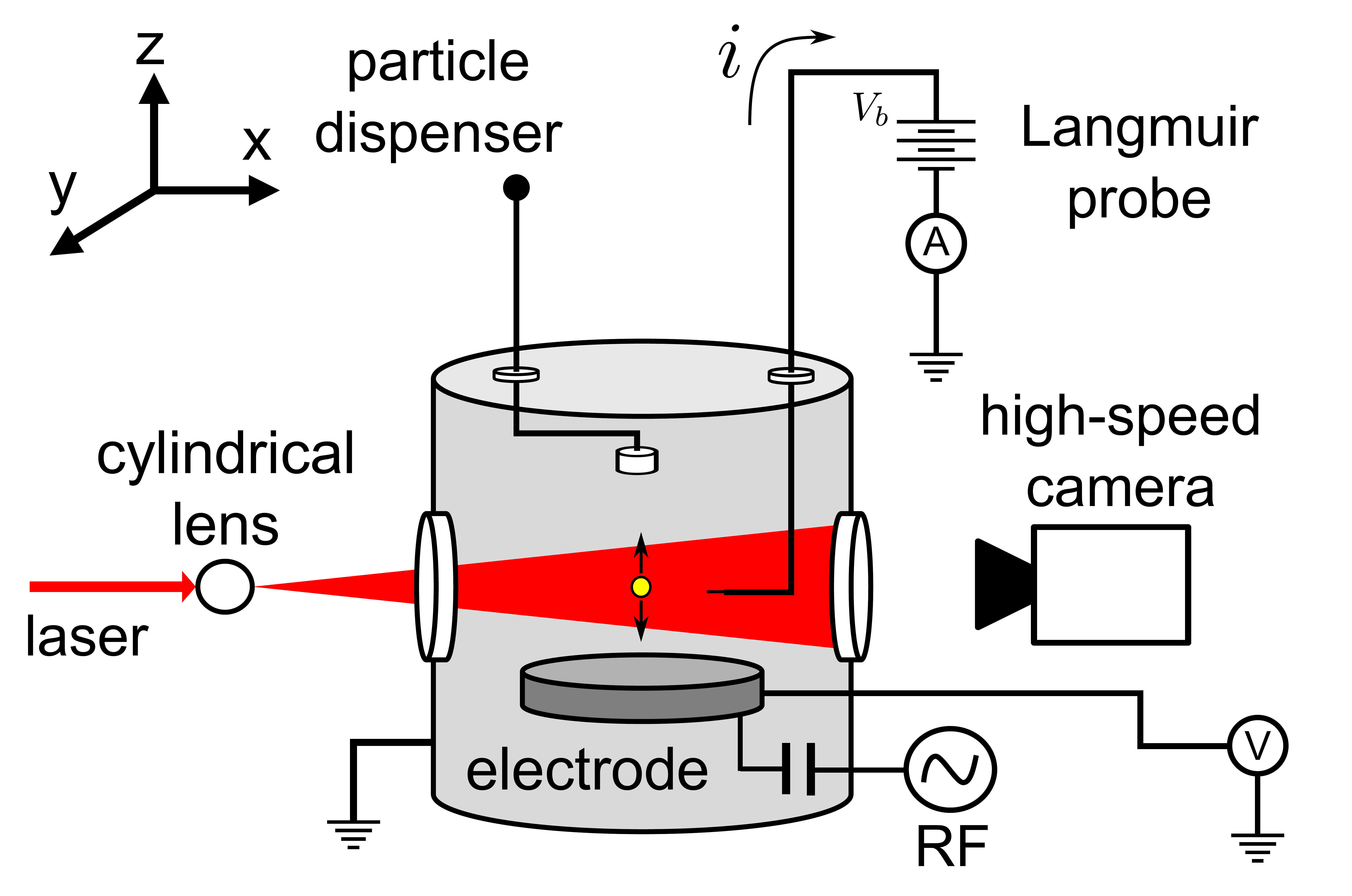}
	\caption{\label{setup} Experimental setup for imaging particle motion \cite{Gogia2017}. The particle levitates in the rf sheath above the electrode and scatters the incoming laser light so it can be imaged with the high-speed camera. The vacuum system has ports so that the particle dispenser and Langmuir probe can be manipulated externally without losing a vacuum.}
\end{figure}

\begin{figure}[!]
	\centering
	\includegraphics[width=\columnwidth]{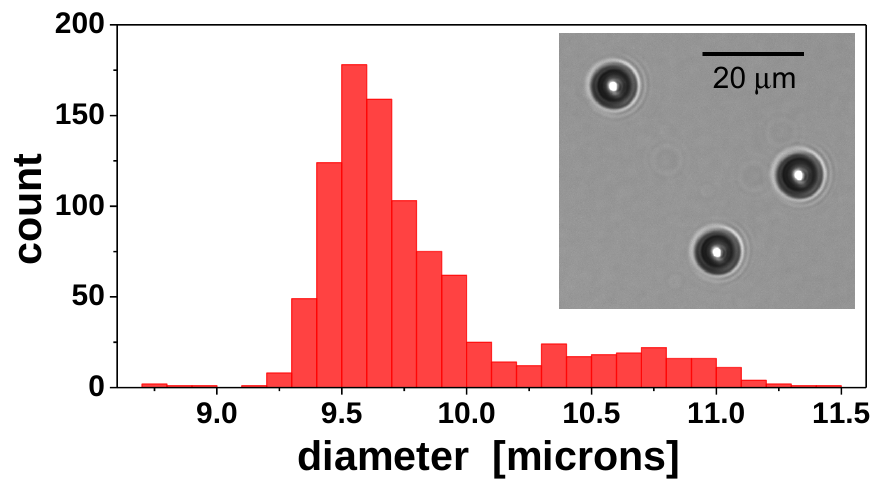}
	\caption{\label{PSD} Typical particle size distribution for melamine-formaldehyde particles with nominal diameters of 9.46 $\mu$m (as reported by the manufacturer). The results were obtained by bright-field optical microscopy, as shown in the inset. The mean of the distribution is 9.83 $\mu m$ and the median is 9.67 $\mu m$. The discrepancy between our measured values and the manufacturer-provided values is due to diffraction-limited imaging of the particles' edge, making the particles appear 2-3\% larger than they actually are.}
\end{figure}

\section{Methods}

\subsection{Experimental setup}

\begin{figure*}[!]
	\centering
	\includegraphics[width=\textwidth]{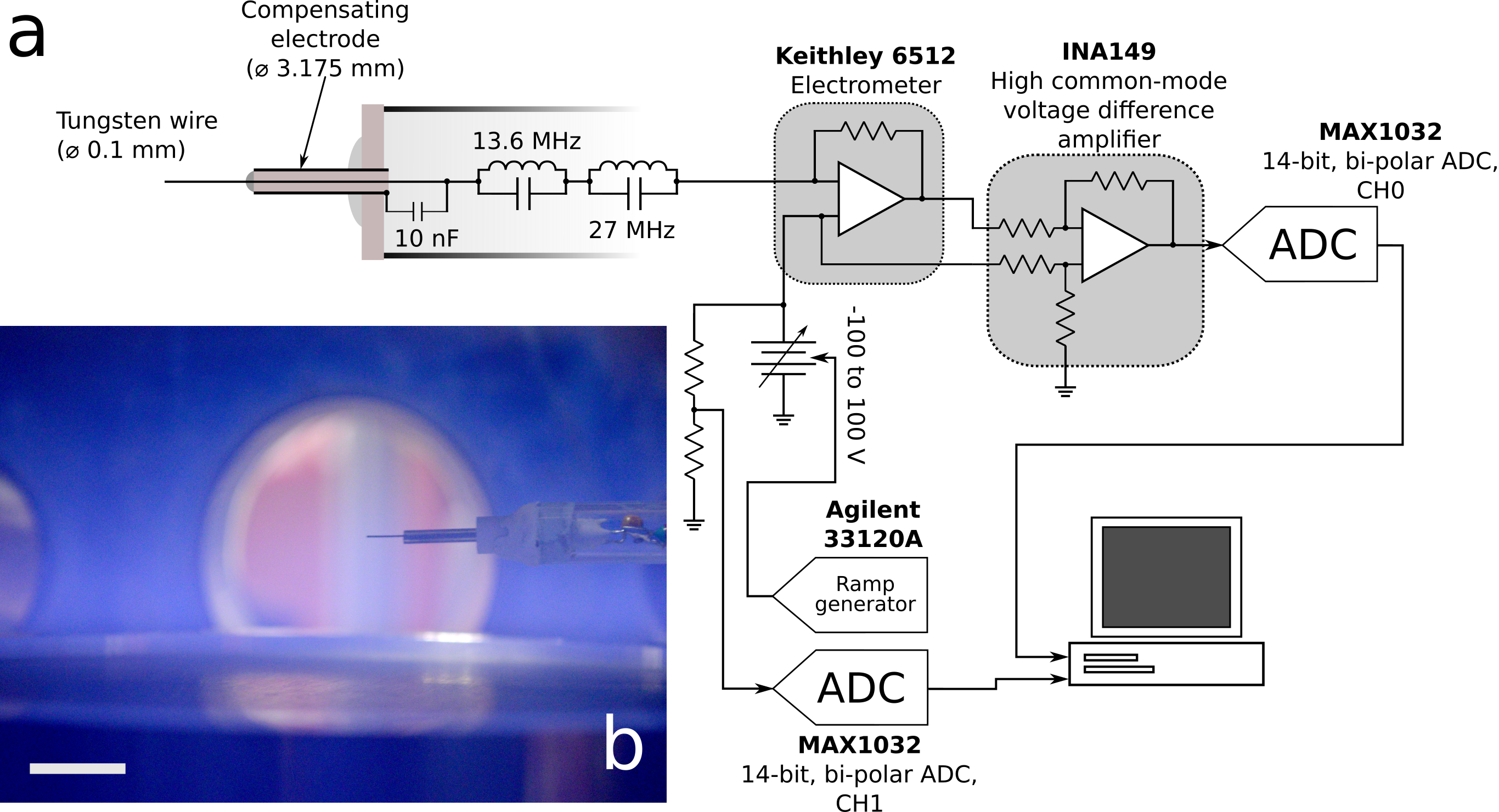}
	\caption{\label{probe} (a) Schematic of the compensated Langmuir probe used to characterize the plasma environment. See text for a detailed description. (b) Photograph of the Langmuir probe near the sheath boundary. The scale bar is 10 mm.}
\end{figure*}

The experiments were carried out in a conventional GEC RF reference cell \cite{Thomas1994,Ivlev2015,Gogia2017} (see Fig.\ \ref{setup} for a simplified schematic rendering). The system consists of a stainless-steal chamber which encloses a weakly-ionized argon plasma and the particles. The plasma was generated by a rf power supply (operating at 13.56 MHz), capacitively-coupled to an aluminum disk (diameter = 15 cm) electrode near the bottom of the chamber. A particle reservoir was suspended over the electrode by a movable arm. The arm passed through the chamber wall, allowing the user to gently shake the reservoir and dispense a small quantity of particles into the plasma. A ring electrode 6 mm in height running along the edge of the disk electrode provided horizontal, electrostatic confinement to the particles. We denote the negative bias developed on the electrode as $\phi_{\text{dc}}$. 

For this work, we used both melamine-formaldehyde (MF) particles with nominal diameters of 8.00, 9.46, and 12.80 $\mu$m, as well as silica particles with nominal diameters of 6.27 $\mu$m. All particles had a coefficient of variance of 1.0\%-1.5\% according to the manufacturer (microParticles GmbH). However, we also characterized the sizes of over 1000 particles using optical microscopy, and analyzed the images in ImageJ \cite{Schneider2012}. A representative histogram for 9.46 $\mu$m MF particles, along with a photograph taken through the microscope, is illustrated in Fig.\ \ref{PSD}. Note that the distribution has a second peak at larger particle sizes that was not reported by the manufacturer, but is consistent with our internal experiments with many particles that show a low population of particles that sit slightly below the main crystalline layer \cite{Gogia2017}. The mean is shifted to larger values than those indicated by the manufacturer, but this is likely due to 1-2 pixels of error in determining the edge of the particle in microscope images. It is noteworthy that results from previous experiments that characterize size by the oscillation frequency do not show a second peak, albeit with limited sample sizes \cite{Carstensen2013}. Particles are also known to shrink in the plasma environment, as characterized by high-resolution, in-situ measurements \cite{Kohlmann2019}.

As mentioned above, both the particles and the electrode acquire net negative surface charges. 
The particles levitate above the disk electrode at the position where the vertical electrostatic force balances the gravitational force. This position typically corresponded to a few millimeters below the edge of the sheath. The thickness of the sheath varied inversely with pressure, and was typically 1-2 cm, as measured by the analysis of the particle motion (see Sec.\ \ref{charge_mod}). Although the drag force from the accelerated ions in the plasma sheath can contribute to this vertical force balance, our estimates of this force in our experimental conditions (see Sec.\ \ref{forces}) suggests that ion drag forces are significantly smaller than either gravity or the electrostatic force. 

The self-induced vertical oscillations are commonly studied in a system consisting of large number of particles. In order to avoid any collective effects, we focused on characterizing the dynamics of a \textit{single levitated particle}. A system comprising an individual suspended particle (see Fig.\ \ref{setup}) was produced by reducing the rf power supply's duty cycle (in essence, pulsing the plasma), causing some of the suspended particles to fall out during the ``off'' portion of the period. The duty cycle was returned to 100$\%$ once all but one particle were removed. We note that this process may preferentially select for smaller particles in the distribution (Fig.\ \ref{PSD}). We verified the presence of a single isolated particle by scanning a laser sheet through the chamber. Two variables were then adjusted to produce vertical oscillations: the gas pressure and the rf power delivered to the gas. We use $\phi_{\text{dc}}$ (as measured by a high-impedance electrometer) as a proxy for the plasma power. Previous studies \cite{Gogia2017} reported the occurrence of vertical oscillations in this experimental setup at pressures $P<$ 1 Pa and bias voltages in the range between -6 and -40 V. Here, we investigate vertical oscillations under similar conditions.

To assess the dynamics of an oscillating particle, we used a high-speed camera (Phantom v7.11, Vision Research). The camera was configured to record at 1000 frames per second, and the resolution of the imaging system was 23 pixels/mm. To visualize the levitated particle, we illuminated it with a 100 mW vertical laser sheet (632 nm) created by passing the beam through a cylindrical lens. 

We employed two approaches to measure the peak-to-peak amplitude of an oscillating particle. The first consisted in sweeping a horizontal laser sheet in the vertical direction from one extremum of the oscillation cycle to the other (i.e from the point the particle was just visible to the point where it just disappeared) and measuring the laser's displacement. Although the laser was mounted onto a micrometer stage, the overall error in this approach was on the order of $\pm$ 1 mm. The amplitudes obtained through this method are rendered in Fig. \ref{amp_vs_pres}b. 

We also tracked the particle's motion using high speed image sequences and the open-source software TrackPy \cite{Allan2019}. This software is capable of sub-pixel accuracy for tracking the center-of-mass of particles in an image sequence. Such analysis not only gave us better estimates of the peak-to-peak amplitude of the oscillation (to an accuracy of $\sim$ 40 $\mu$m), but also allowed us to extract the velocity profile of the motion. Figure \ref{crystal_gas}c shows two cycles of an oscillation characterize by this method while Fig. \ref{amp_vs_pres}c shows oscillation amplitudes extracted through this method (discussed further in Sec. IIIA).

\begin{figure}[!]
	\centering
	\includegraphics[width=\columnwidth]{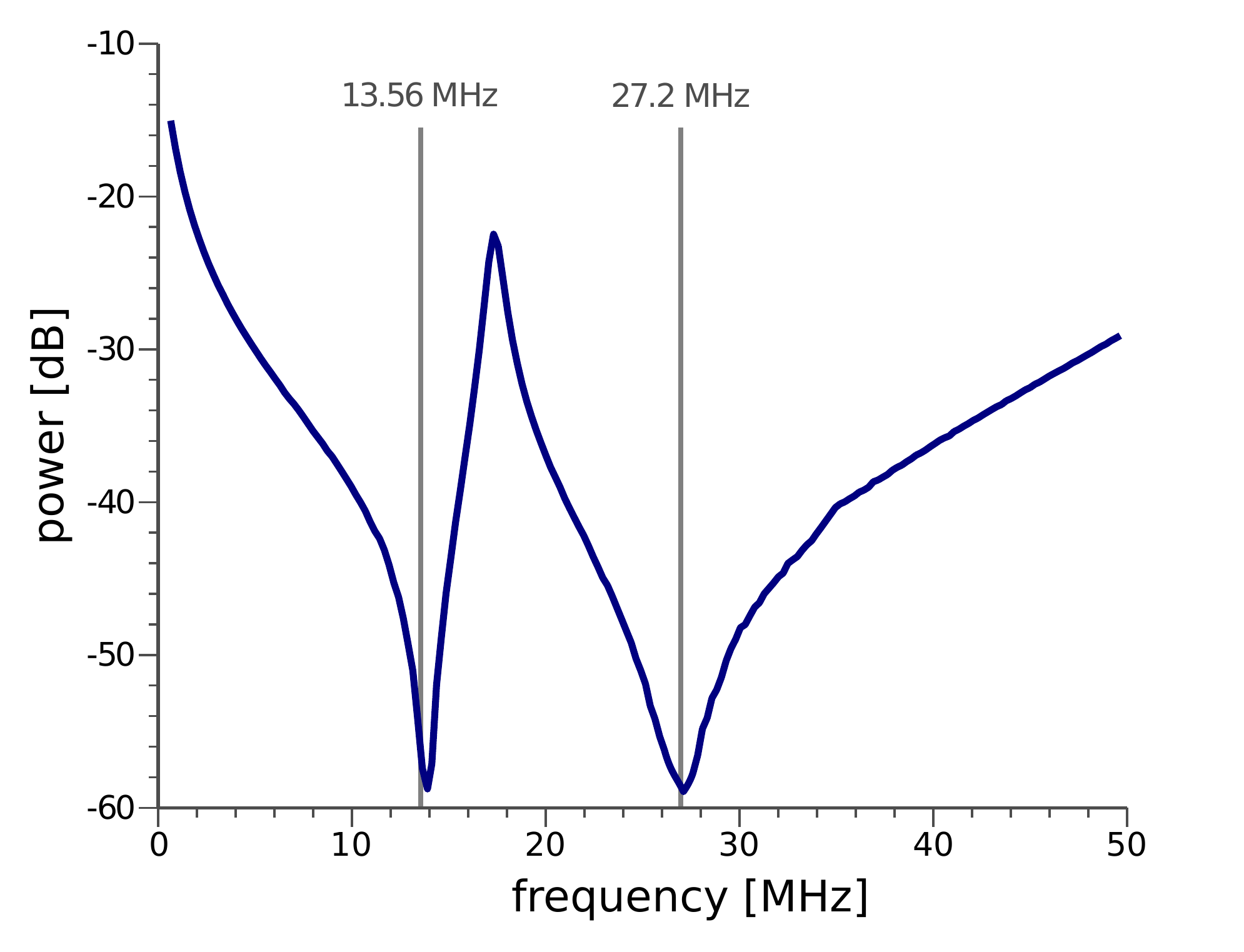}
	\caption{\label{response} Frequency response of the Langmuir probe, as measured with a network analyzer, showing high impedance near 13.56 MHz and 27.2 MHz.}
\end{figure}  

\subsection{Langmuir Probe}

In addition to our high-speed imaging setup, we developed a custom, compensated Langmuir probe to characterize the plasma in which suspended particles oscillated. The probe is described schematically in  Fig.\ \ref{setup}. We note that all probe measurements were conducted with the particles \textit{absent} from the system to ensure we were characterizing only plasma parameters. In its most basic form, a Langmuir probe consists of a thin wire with radius $r_\text{P}$ inserted into the plasma and then biased to some potential $\phi_\text{b}$ relative to a reference node (here, the grounded chamber wall). The potential difference between the biased probe and plasma produces a sheath around the probe, resulting in a current flow through the probe which carries information regarding the plasma environment. Our probe design makes use of ``low pressure theory,'' which implicitly assumes that $r_\text{P}$ is much smaller than the characteristic Debye length $\lambda_\text{D}$ and $\lambda_\text{D} << \lambda_{\text{en}}$, where $\lambda_{\text{en}}$ is the electron-neutral mean free path \cite{Godyak1992, ryan2019comparison}. For the plasma system described above, the electron Debye length $\lambda_\text{D}\approx$ 1-2 mm, and $\lambda_{\text{en}}$ falls between the range $1 - 5$ cm in the bulk plasma (see Sec.\ \ref{plasma}). 

The theory of Langmuir probes has been addressed in great detail in other works \cite{Godyak1992, Demidov2002, Chen2003, Merlino2007}, however it is worthwhile to briefly summarize some key principles. When the probe has sufficient negative bias in relation to the plasma, a sheath around the probe effectively repels electrons and the current to the probe is the result of ions that random-walk past the sheath boundary.  As the probe potential is made more positive, the ratio of electrons to ions collected by the probe increases, generating an electron retardation current. At some probe potential $\phi_\text{f}$ the number of ions arriving at the probe equals the number of electrons and the current through the probe goes to zero. This \textit{floating potential} is characteristic of the equilibrium charge gained by objects immersed in the plasma, such as particles. Above the ion saturation current, the electron current through the probe grows exponentially because of the exponential form of the Maxwell-Boltzmann electron velocity distribution in the plasma:

\begin{equation} \label{eq1}
I_\text{e} (\phi_\text{b}) = I_\text{es} \exp{\dfrac{e [\phi_\text{b} - \phi_\text{p}]}{k_\text{B} T_\text{e}}}.
\end{equation}

In Eq.\ \ref{eq1}, $e$ is the elementary charge, $T_\text{e}$ is the electron temperature, and $k_\text{B}$ is Boltzmann's constant. However, above the plasma potential $\phi_\text{p}$ the current no longer increases at an exponential rate. Here, current increase is solely due to an expanding collection region around the probe. At probe voltages larger than $\phi_\text{p}$, the electron saturation current $I_\text{es}$ is reached and is given by: 
\begin{equation} \label{eq2}
I_\text{es} = e n_\text{e} A  \bigg( \dfrac{k_\text{B}T_\text{e}}{2 \pi m_\text{e}} \bigg)^{1/2}
\end{equation}
where $A$ is the exposed area of the probe, $n_\text{e}$ is the electron number density, and $m_\text{e}$ is the electron mass. 

Physically, the probe is a cylindrical tungsten wire ($r_\text{P}$ = 50 $\mu$m) housed within a borosilicate glass tube (outer diameter 6.35 mm). Only a length of 5 mm of the wire is exposed horizontally to the plasma (see Fig.\ \ref{probe}). All vacuum seals were made with TorrSeal epoxy. Because the plasma is generated by an rf source, the plasma parameters vary in conjunction with the source. Thus, we compensated our probe based on the design of Chen \cite{Chen2009}. In essence, compensation involves forcing the probe to follow the plasma's AC component so that the voltage drop across the probe sheath ($\phi_\text{b} - \phi_\text{p}$) remains constant. Thus, we increased the probe's impedance to ground at 13.56 MHz and the first harmonic (27.2 MHz) using two resonant tank circuits in series. Additionally, an auxiliary electrode, capacitively-coupled to the probe tip, was required to ensure that the rf sheath impedance is lower than the impedance to ground. The auxiliary electrode has a much larger area than the probe tip and was placed coaxially with the probe tip. The frequency response of the compensated probe is shown in Fig. \ref{response}, showing anti-resonances at 13.56 and 27.2 MHz. 

\section{Experimental Results}

\subsection{Description of single-particle oscillations}

\begin{figure*}[!] 
	\centering
	\includegraphics[width=6.5 in]{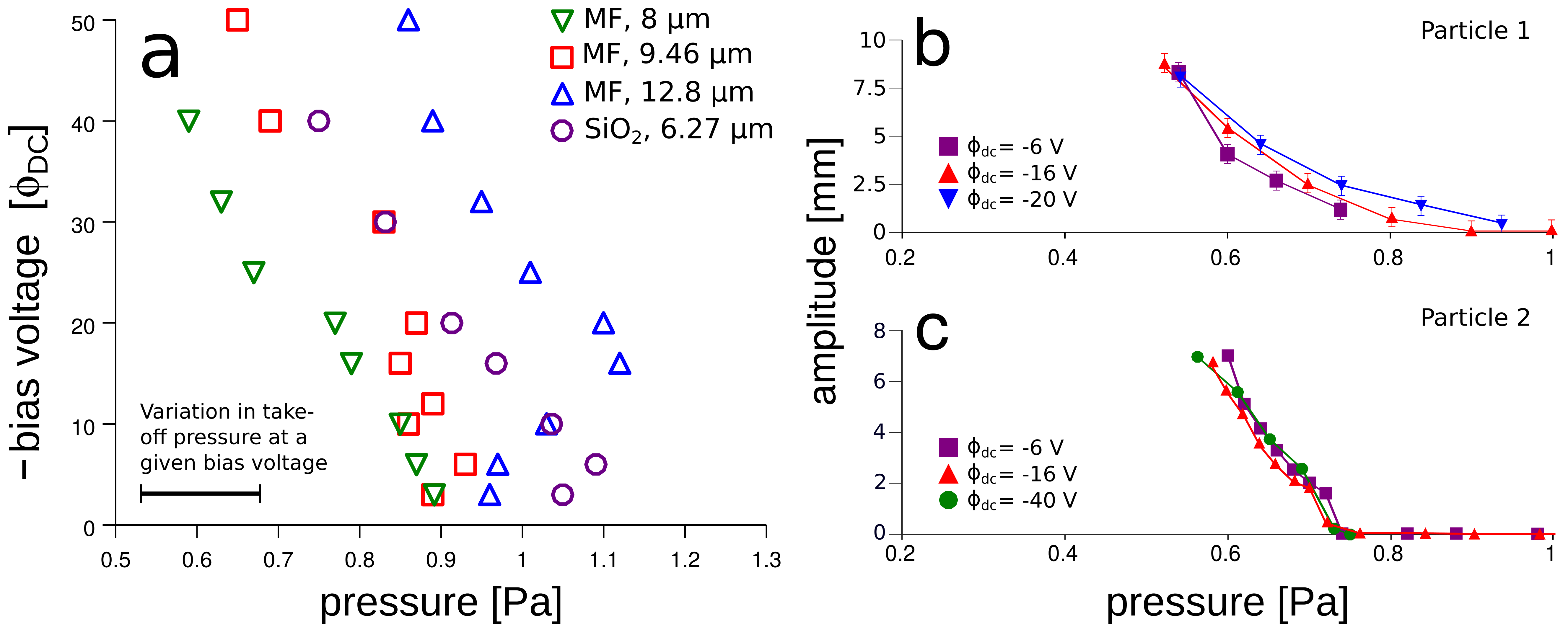}
	\caption{\label{amp_vs_pres} (a) The pressure/bias voltage combinations under which particles of different sizes and material compositions begin to oscillate. The uncertainty in pressure at which particles from a given sample begin to oscillate is described by the black horizontal error bar. (b) The amplitude of oscillation as a function of pressure and bias voltage for an MF particle with a nominal diameter of 9.46 $\mu$m. This data was acquired with help of a mechanical actuator resulting an uncertainty in the amplitude of 1 mm (note error bars). (c) The same measurement as in (b) but with a different MF particle of the same nominal diameter. For this data, the particle's amplitude was measured using a high-speed camera and a particle image velocimetry algorithm (TrackPy). Thus, the error is much smaller than in (b), on the order of a 40 microns. Despite the different measurement techniques, there is good agreement between (b) and (c).}
\end{figure*}

\begin{figure}[!]
	\centering
	\includegraphics[width=3.0 in]{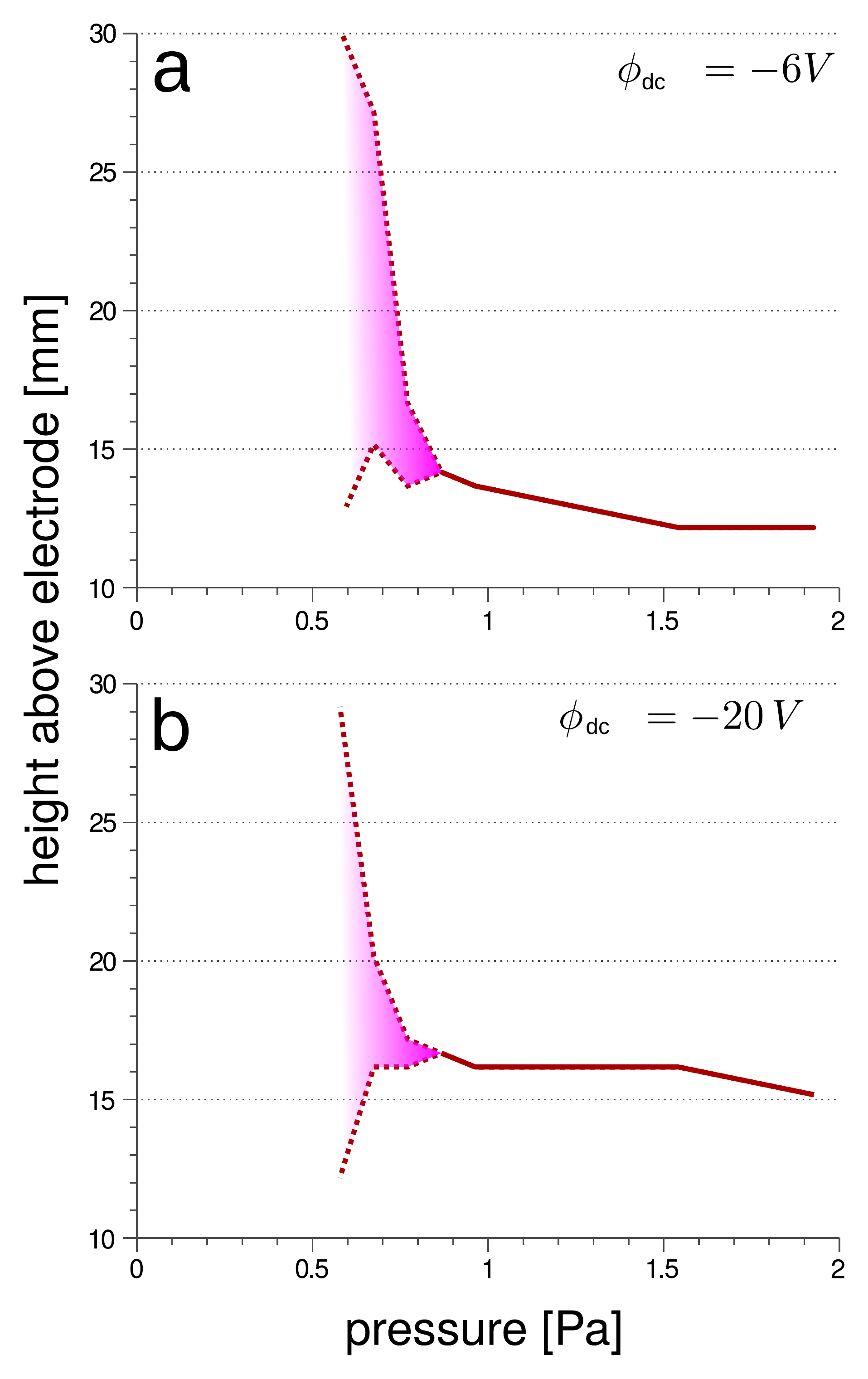}
	\caption{\label{amp_peak_peak} Position of an MF particle (9.46 $\mu$m diameter) as a function of pressure for two different bias voltages $\phi_\text{dc}$: (a) -6 and (b) -20 V. In the two panels, the shaded areas denote the pressures at which particles oscillate. The upper and lower dotted lines show the lower and upper limits of the oscillation. Here, the uncertainty in particle position is $\pm$ 1 mm.}
\end{figure} 

At elevated gas pressures and plasma powers, a particle suspended in the plasma remains stationary at the point where the electrostatic force balances that of gravity. As mentioned previously, once the pressure is reduced below some threshold, the particle undergoes spontaneous vertical oscillations.  However, we found that the precise value of the pressure threshold for a given particle depends on the bias voltage on the electrode. The combinations of plasma powers and gas pressures at which oscillations commence for different particle compositions and sizes are shown in Fig.\ \ref{amp_vs_pres}a. Notice that as $|\phi_\text{dc}|$ (and plasma power) is reduced, the onset of the vertical oscillations occurs at higher gas pressures. The range of pressures at which the particles oscillate is narrower than the range of bias voltages, implying that the oscillations are more strongly dependent on $P$ than on $\phi_\text{dc}$. 

To illustrate this point in greater detail, we measured  the oscillation amplitude for two different 9.46 $\mu$m MF particles as a function of both $P$ and $\phi_{\text{dc}}$ (Fig.\ \ref{amp_vs_pres}b-c) through two different methods (as discussed in Sec. IIA and caption of Fig.\ \ref{amp_vs_pres}).  Although data was extremely repeatable for a given particle, there was significant variation between particles obtained from the same sample, presumably due to possible size variations (Fig.\ \ref{PSD}). For a given bias voltage, the transition between the stationary and oscillatory regimes occupies a very narrow pressure range. Indeed, by lowering the plasma pressure by only a few tenths of a Pascal, the particle oscillation amplitude increases from 0 mm to 10 mm, a size comparable to the extent of the plasma sheath. 

\begin{figure}[!]
	\centering
	\includegraphics[width=\columnwidth]{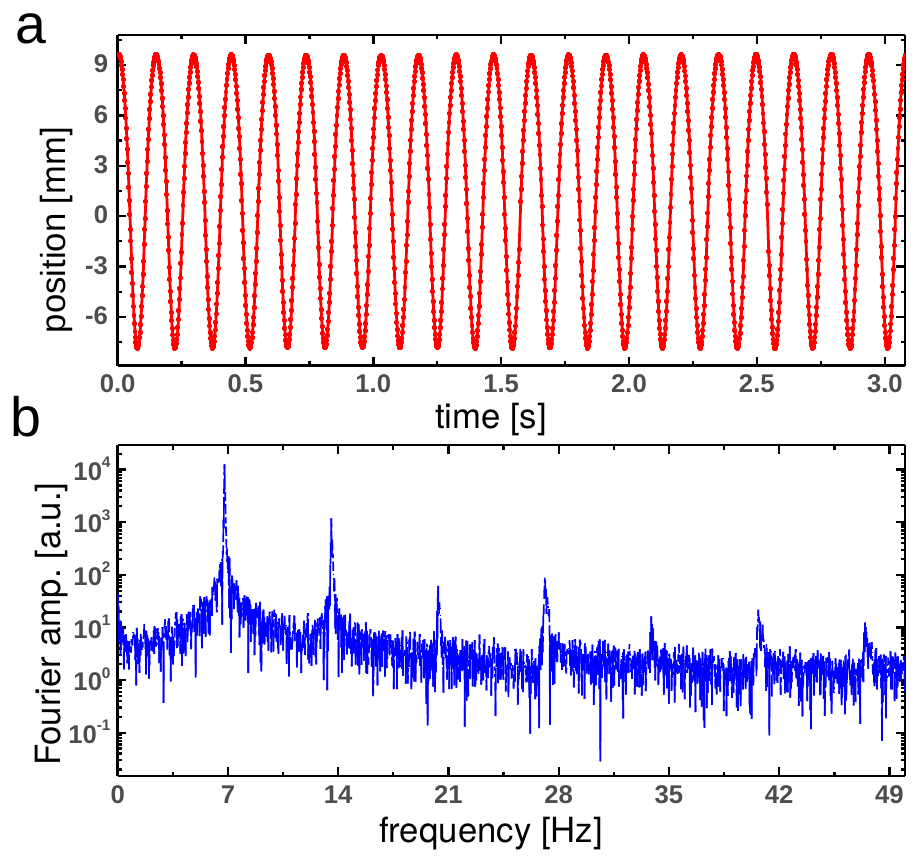}
	\caption{\label{fourier} (a) Vertical position vs. time for a single MF particle (diameter = 9.46 $\mu$m). The constant-amplitude oscillation persists for minutes. (b) Fourier power spectrum of the particle oscillation. Both even and odd harmonics are visible.}
\end{figure}

While reports of spontaneous vertical oscillations of particles in plasmas are copious in the literature \cite{Nunomura1999,Ivlev2000Inf,Samarian2001,Vaulina2001,Resendes2002,Sorasio2002}, the oscillations we observe in our experiments differ in two fundamental ways. First, the amplitudes of the oscillations are more than 10 times larger than those reported in previous studies. At very low pressures and plasma powers, particles in our system undergo vertical excursions of several millimeters, even exceeding 1 cm in some cases, whereas previous experiments produced oscillations not much greater than several hundred microns. Also, the particle oscillation is particularly asymmetric at large amplitudes. Figure \ref{amp_peak_peak} shows the maximum and minimum positions of a MF particle as a function of pressure for two different bias voltages. As the pressure is lowered, the Debye screening length and the equilibrium position of the particle both increase. The strong asymmetry of the motion about the equilibrium position reflects the highly nonlinear variation of the electrostatic force throughout the cycle. As we will show in Sec.\ \ref{charge_mod}, the particle actually exits the sheath and spends a significant portion of its cycle in free fall.

Second, the single-particle oscillation displays a regularity in both amplitude and frequency not seen in previous works. A time series of the vertical position of a single MF particle (9.46 $\mu$m) is shown in Fig.\ \ref{fourier}a. The regularity in the amplitude is striking and it persists for minutes. As will be discussed, this regularity strongly suggests that stochastic variations in the charge or plasma environment are not responsible for driving the particle motion. Additional information can be gained from analyzing the spectral content of the particle's trajectory. A power spectrum of the particle motion is shown in Fig.\ \ref{fourier}b. Note that, while themost of the energy is concentrated in the $\sim$ 7 Hz fundamental, the oscillations display activity across odd and even harmonics. In other words, the motion of the particle is not strictly sinusoidal, but rather exhibits considerable anharmonicity.

\subsection{Plasma characteristics} \label{plasma}

\begin{figure}[!]
	\centering
	\includegraphics[width=3.2 in]{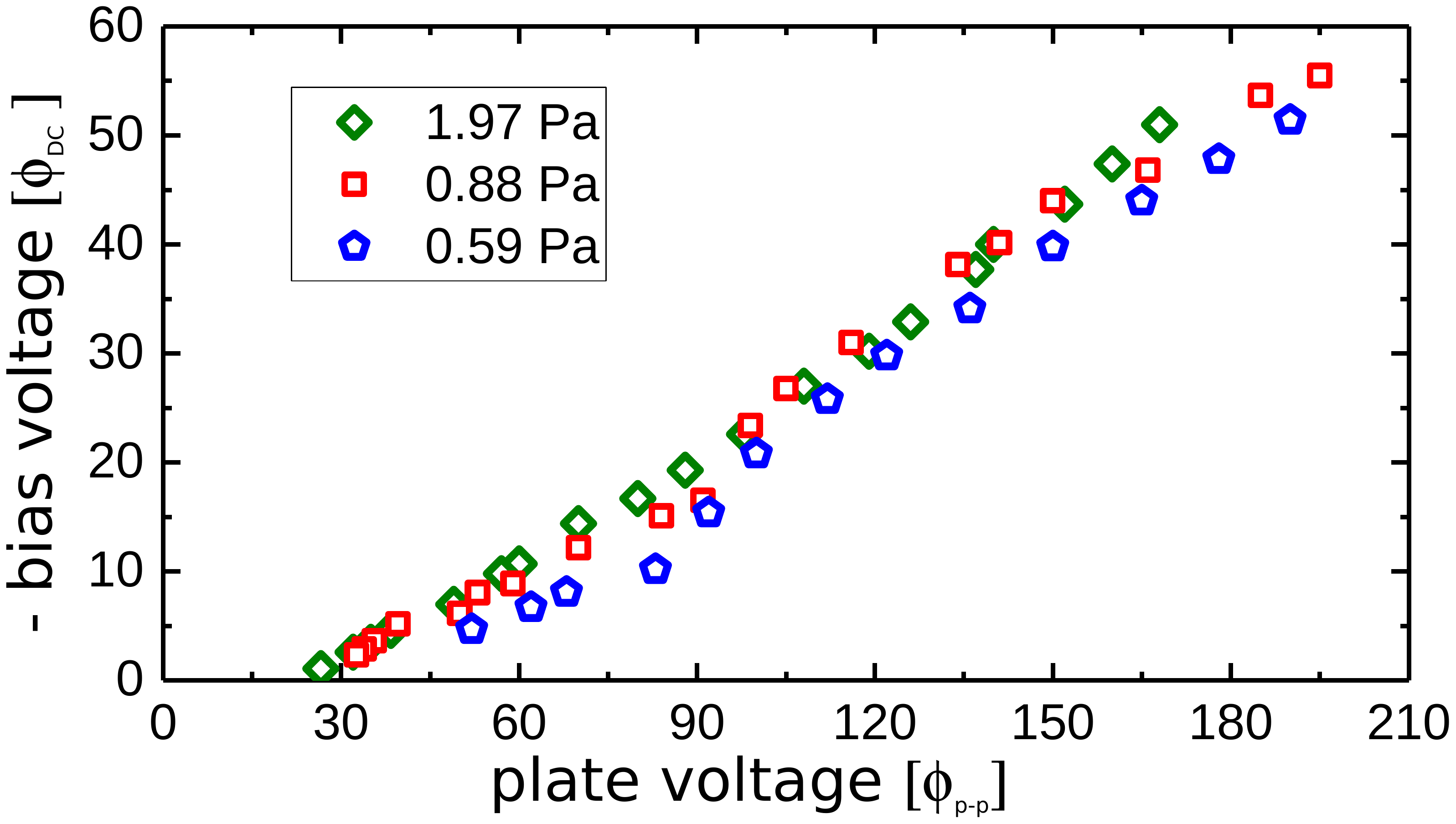}
	\caption{\label{mean_peak} The dc bias voltage on the electrode $\phi_\text{dc}$ as a function of the peak-to-peak ac voltage applied to the electrode at 13.56 MHz. Data is shown for three different values of the pressure. At lower pressures, the data deviates more significantly from linear behavior.}
\end{figure}

Before analyzing the motion of suspended particles, we discuss the plasma environment in which they levitate. As aforementioned, the plasma was generated through a powered disk electrode (see Fig.\ \ref{setup}). While we use the bias voltage $\phi_\text{dc}$ as a proxy for the plasma intensity, we refer the reader to Fig.\ \ref{mean_peak} which shows $\phi_\text{dc}$ as a function of the peak-to-peak ac voltage applied to the electrode for three representative pressures. The behavior is mostly linear, and there is a relatively weak dependence of pressure. Unless otherwise noted, all reported voltages are reported relative to system ground. 

\begin{figure}[!]
	\centering
	\includegraphics[width=3.2 in]{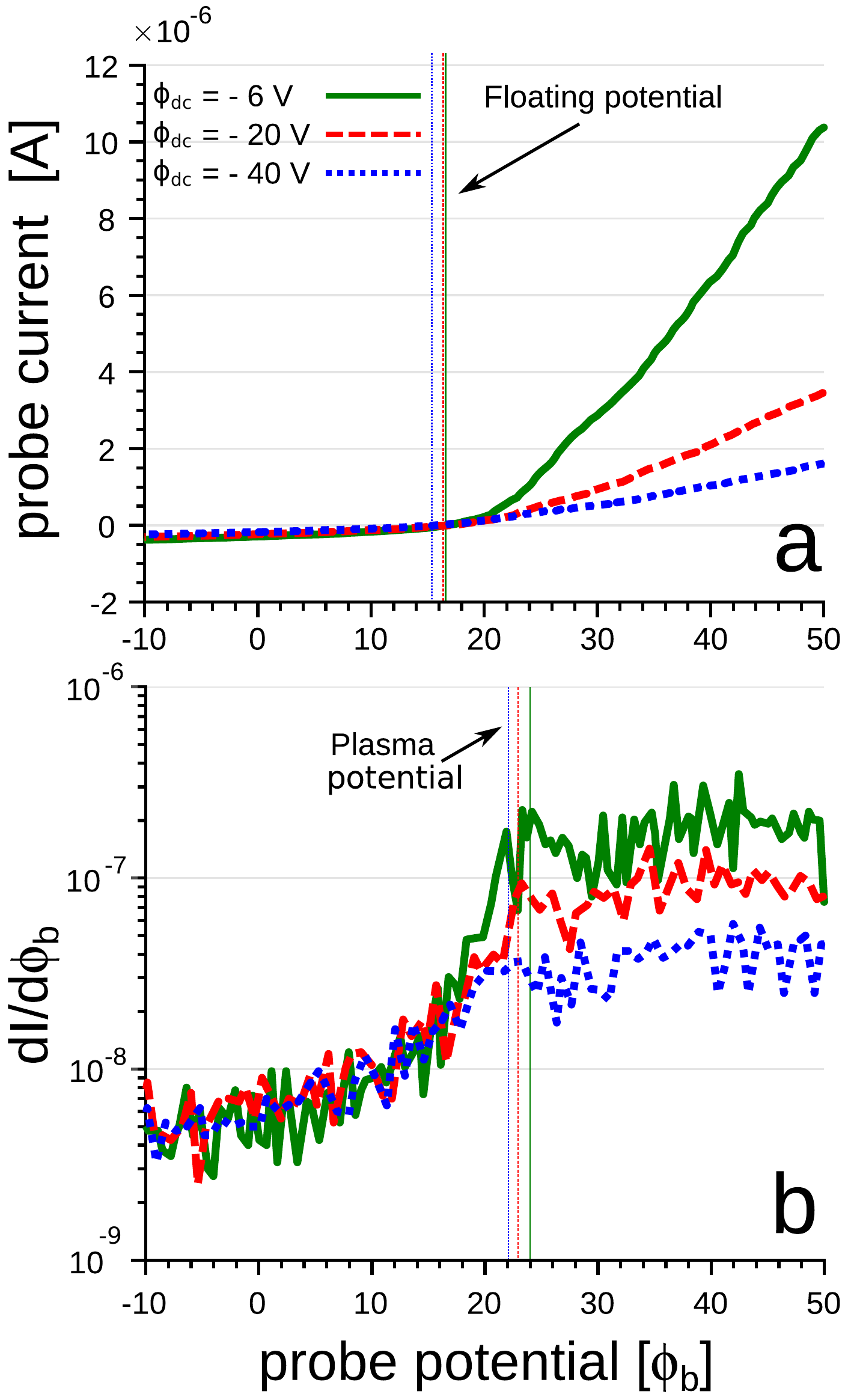}
	\caption{\label{probe_current} (a) Langmuir probe measurements for the collection current (I-$\phi_\text{b}$) as a function of probe potential for three different values of $\phi_{\text{dc}}$ at $P=1.0$ Pa (a pressure corresponding to the threshold at which oscillations begin). The height of the probe above the rf electrode was 40 mm (in the bulk plasma). The point at which the current crosses zero in (a) indicates the floating potential $\phi_\text{f}$. Here, the compound uncertainty in the current measurement is $\pm$ 10 pA and the uncertainty in $\phi_\text{b}$ is $\pm$ 150 mV. (b) The derivatives of the I-$\phi_\text{b}$ characteristics in (b). The change in slope in is representative of the plasma potential $\phi_\text{p}$. The error in the estimate of the plasma potential is $\pm$ 1-2V.} 
\end{figure}

We used a Langmuir probe to characterize basic plasma properties such as the floating potential $\phi_\text{f}$, the plasma potential $\phi_\text{p}$, and ion number densities. The current-voltage characteristics for our Langmuir probe are shown in Fig.\ \ref{probe_current}a for a pressure of 1.02 Pa and $\phi_{\text{dc}}$ = -6, -20, and -40 V. The voltage on the probe $\phi_\text{b}$ was swept between -100 and 100 V relative to system ground (for clarity, only the response between -10 to 50 V is shown in Fig.\ \ref{probe_current}). The dashed vertical lines indicate the points where the current in the Langmuir probe was 0 A, representative of the floating potential $\phi_\text{f}$ for each of the three electrode voltages. For $\phi_{\text{dc}}$ = -6 V, $\phi_\text{f}$ $\approx$ 15.5 V; for $\phi_{\text{dc}}$ = -20 V $\phi_\text{f}$ $\approx$  16.5 V; and, for $\phi_{\text{dc}}$ = -40 V $\phi_\text{f}$ $\approx$ 16.5 V. We note that over the 200 V-span of $\phi_\text{b}$ there is an uncertainty of  $\pm$ 150 mV. The compound uncertainty in the current measurement is $\pm$ 10 pA. For much higher values of $\phi_\text{b}$, the probe emitted electrons, as indicated by a characteristic visible glow around the probe tip.

As mentioned above, when the voltage applied to the Langmuir probe is raised above a threshold $\phi_\text{p}$, the current no longer increases exponentially and increases slowly as the collection region around the probe expands. This \textit{plasma potential} can be found by differentiating the I-$\phi_\text{b}$ characteristics in Fig.\ \ref{probe_current}a and locating a change in slope (Fig.\ \ref{probe_current}b). For $\phi_{\text{dc}}$ = -6 V, $\phi_\text{p}$ $\approx$ 22 V; for $\phi_{\text{dc}}$ = -20 V, $\phi_\text{p}$ $\approx$ 23 V, and, for $\phi_{\text{dc}}$ = -40 V, $\phi_\text{p}$ $\approx$ 24 V. Due to differentiation, the error in the estimate of the plasma potential is $pm$ 1-2V. The floating potential is defined at the point where $I_{\text{ion}}$ = $I_\text{e}$, where $I_\text{e}$ is given by Eq.\ \ref{eq1}. For the current study, we estimate $I\textsubscript{ion}$ from the Bohm current \cite{Chen2003}:
\begin{equation} \label{eq3}
I_\text{ion} \approx \dfrac{n_\text{i} e A}{2} (k_\text{B} T_\text{e}/m_\text{i})^{1/2}, 
\end{equation}
where $m_\text{i}$ is the ion mass, and $n_i$ is the ion number density. Equating Eq.\ \ref{eq1} and \ref{eq3} and assuming quasi-neutrality ($n_\text{e} \approx n_\text{i}$) leads to:
\begin{equation} \label{eq4}
\phi_\text{f} = \phi_\text{p} - \dfrac{k_\text{B} T_\text{e}}{2e} \ln \bigg(\dfrac{2m_\text{i}}{\pi m_\text{e}} \bigg).
\end{equation}
\begin{figure}[!]
	\centering
	\includegraphics[width=3.2 in]{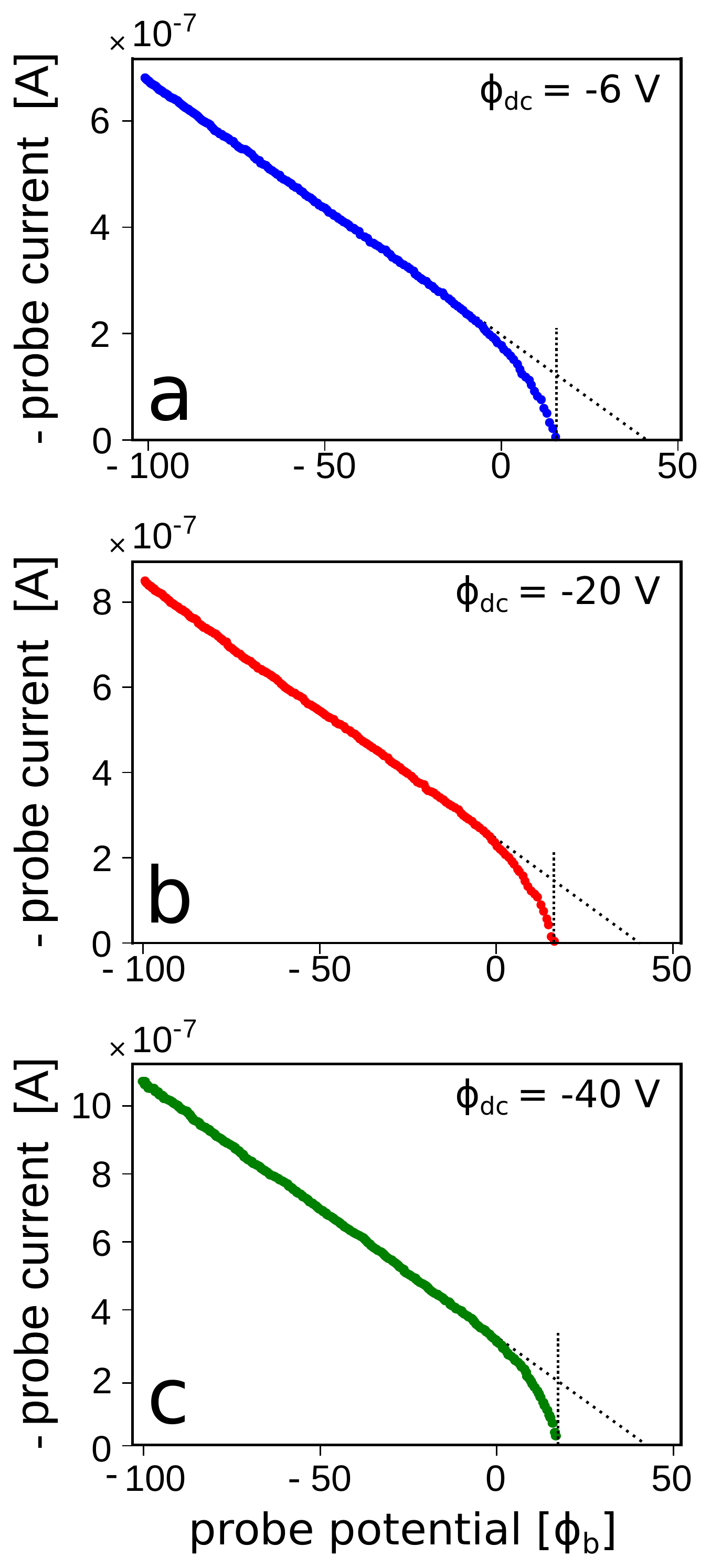}
	\caption{\label{ion_saturation} Zoom-in of the ion current portions of the plasma I-$\phi_\text{b}$ characteristics shown in Fig.\ \ref{probe_current}.  Note that these curves have been plotted with sign reversal. Using the floating potential methodthe described by Chen et al \cite{Chen2002}, we can estimate the ion density $n_\text{i}$ through an extrapolation to the floating  potential  of the ion saturation current (black dashed curves). We perform this analysis for bias potentials of (a) -6 V, (b) -20 V, and (c) -40 V. The ion density at $\phi_\text{dc}$ = -6 V is 2.24 $\times 10^{13}$ $\text{m}^{-3}$; at $\phi_\text{dc}$ = -20 V, 4.28 $\times 10^{13}$ $\text{m}^{-3}$; and at $\phi_\text{dc}$ = -40 V, 5.57 $\times 10^{13}$ $\text{m}^{-3}$}.
\end{figure}
For argon, the above equation can be simplified to: $\phi_\text{f} - \phi_\text{p}\approx -5 k_\text{B} T_\text{e}/e$ (which includes a geometrical correction factor for a cylindrical probe tip \cite{Chen2003}). Thus, for our experimental conditions, $T_\text{e}$ is in the range of 1.3 - 1.5 eV. 

The ion number density $n_\text{i}$ can also be extracted from the I-$\phi_\text{b}$ characteristics using the floating potential method described in Chen et al. \cite{Chen2002}. This method consists in plotting $I^\beta$ v. $\phi_\text{b}$, where $\beta$ is an arbitrary constant, and performing a linear fit. The fit to the ion saturation current is then extrapolated to the floating potential, as shown in Fig.\ \ref{ion_saturation}. Once this quantity is known, $n_\text{i}$ can be computed from the definition of the Bohm current.  While the method, as noted by Chen and company, is fundamentally heuristic, the strategy provides estimates of $n_\text{i}$ that agree well with those obtained through other techniques down to pressures of $\sim$ 0.3 Pa. Given the linearity of the data, we assume that $\beta$ = 1, and the resulting ion density at $\phi_\text{dc}$ = -6 V is 2.24$\times 10^{13}$ $\text{m}^{-3}$; at $\phi_\text{dc}$ = -20 V, 4.28$\times 10^{13}$ $\text{m}^{-3}$; and at $\phi_\text{dc}$ = -40 V, 5.57$\times 10^{13}$ $\text{m}^{-3}$. With this information, the electron Debye length in the plasma system can be calculated as follows \cite{Wiesemann2014}:
\begin{equation} \label{eq4.1}
\lambda_\text{D} = \left( \frac{\epsilon_0 k_\text{B} T_\text{e}}{e^2 n_\text{e}} \right) ^{0.5}.
\end{equation}
Here, $\epsilon_0$ is the permittivity of free space. At $\phi_\text{dc}$ = -6 V, $\lambda_\text{D}$ = 1.78 mm; at $\phi_\text{dc}$ = -20 V, $\lambda_\text{D}$ = 1.39 mm; and at $\phi_\text{dc}$ = -40 V, $\lambda_\text{D}$ = 1.21 mm. 

For ion-ion collisions, the mean free path  $\lambda_\text{ii}$ can be computed using
\begin{equation} \label{eq4.0}
\lambda_\text{ii} = \left(\pi d_\text{c}^2 n_\text{i}\right)^{-1},
\end{equation}
where $d_\text{c} \approx e^2/(4\pi\epsilon_0 m_\text{i}v_\text{i}^2)$ is the argon ion's interaction length enhanced by Coulomb forces. The mean thermal speed $v_\text{i}$ for the ions is given by $\sqrt{8k_\text{B}T_\text{i}/(m_\text{i} \pi)}$. Note that for our experimental conditions the ion temperature is close to the neutral gas temperature, $\sim$ 300 K. Consequently, $\lambda_\text{ii}$ ranges between $\sim$ 30 m (at $\phi_\text{dc}$ = -6 V) to $\sim$ 13 m (at $\phi_\text{dc}$ = -40 V). However, in the sheath, ions lose energy predominantly by collisions with neutrals, which have a much higher density \cite{Khrapak2019}. The mean free path for ion-neutral collisions can be estimated using \cite{Lieberman}
\begin{equation}
\lambda_\text{in} = \dfrac{4.04\times 10^{-3} \text{ Pa m}}{P},
\label{inmf}
\end{equation}
%
%
so that a pressure of 1 Pa corresponds to $\lambda_\text{in}$ = 4.04 mm. As we will show in Sec.\ \ref{charge_mod}, the collisionality of ions can have a non-negligible effect on the sheath's electric field. Lastly, the maximum electron collisional cross-section for the pressure and electron energy regimes in our experiments is $\approx 2 \times 10^{-19}$ m$^{2}$ \cite{ryan2019comparison}. This results in electron-neutral mean free paths $\lambda_\text{en}$ in the range of 1-5 cm. 
Thus, as discussed previously, $\lambda_\text{en}$ is much larger than $\lambda_{\text{D}}$ and $\text{r}_p$, and the use of ``low pressure theory'' is a reasonable approximation for our Langmuir probe experiments. 

\begin{figure}[!]
	\centering
	\includegraphics[width=0.9\columnwidth]{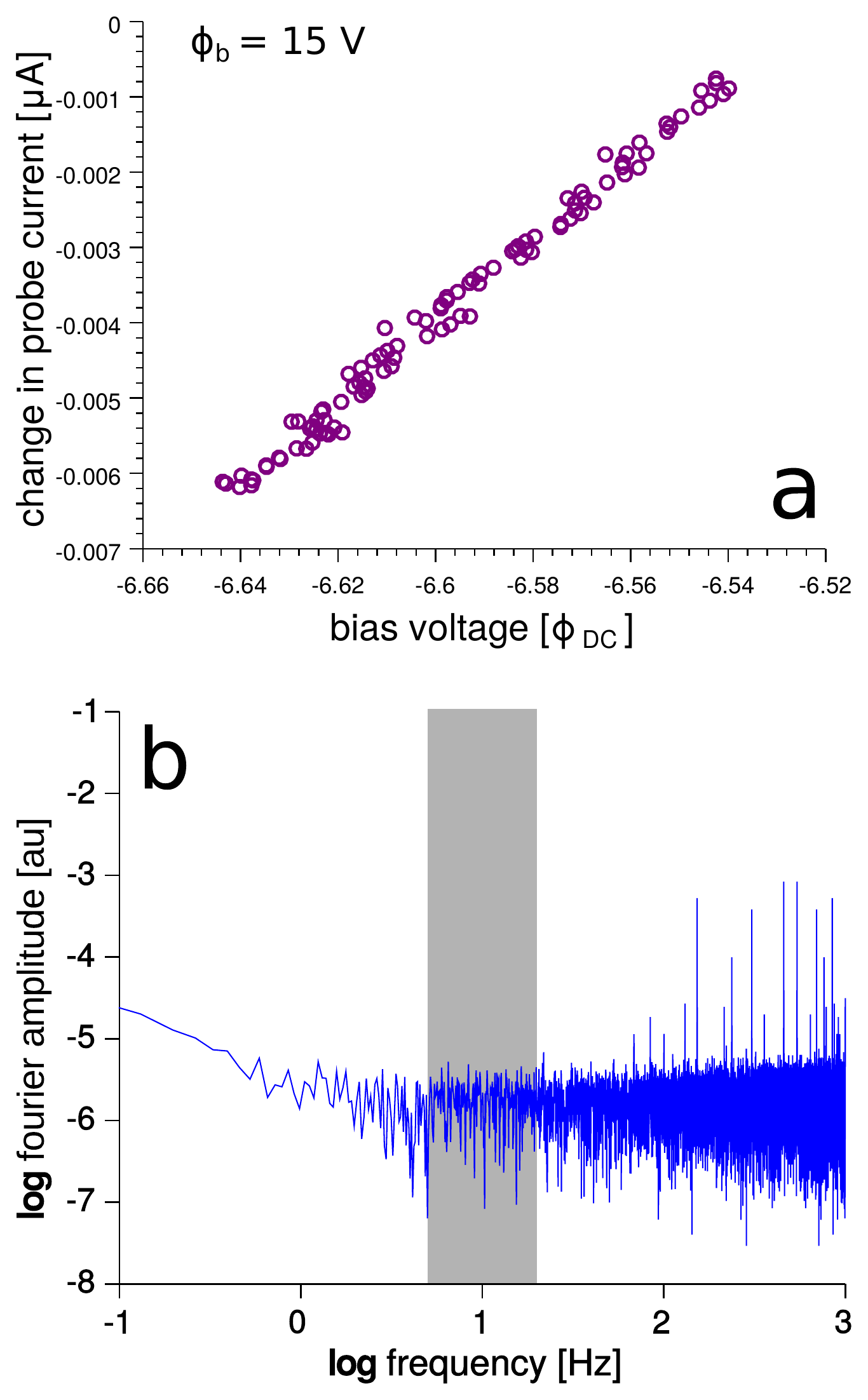}
	\caption{\label{fourier_plasma} (a) Variation in probe current as a function of a small, controlled change in bias voltage. (b) Fourier transform of voltage fluctuations from the Langmuir probe at 15 mm above the electrode (near the sheath boundary). The probe was biased to zero Volts relative to system ground, and the pressure was 0.6 Pa. The gray region depicts the typical frequencies of particle oscillations.}
	
\end{figure}  

\begin{figure}[!]
	\centering
	\includegraphics[width=0.9\columnwidth]{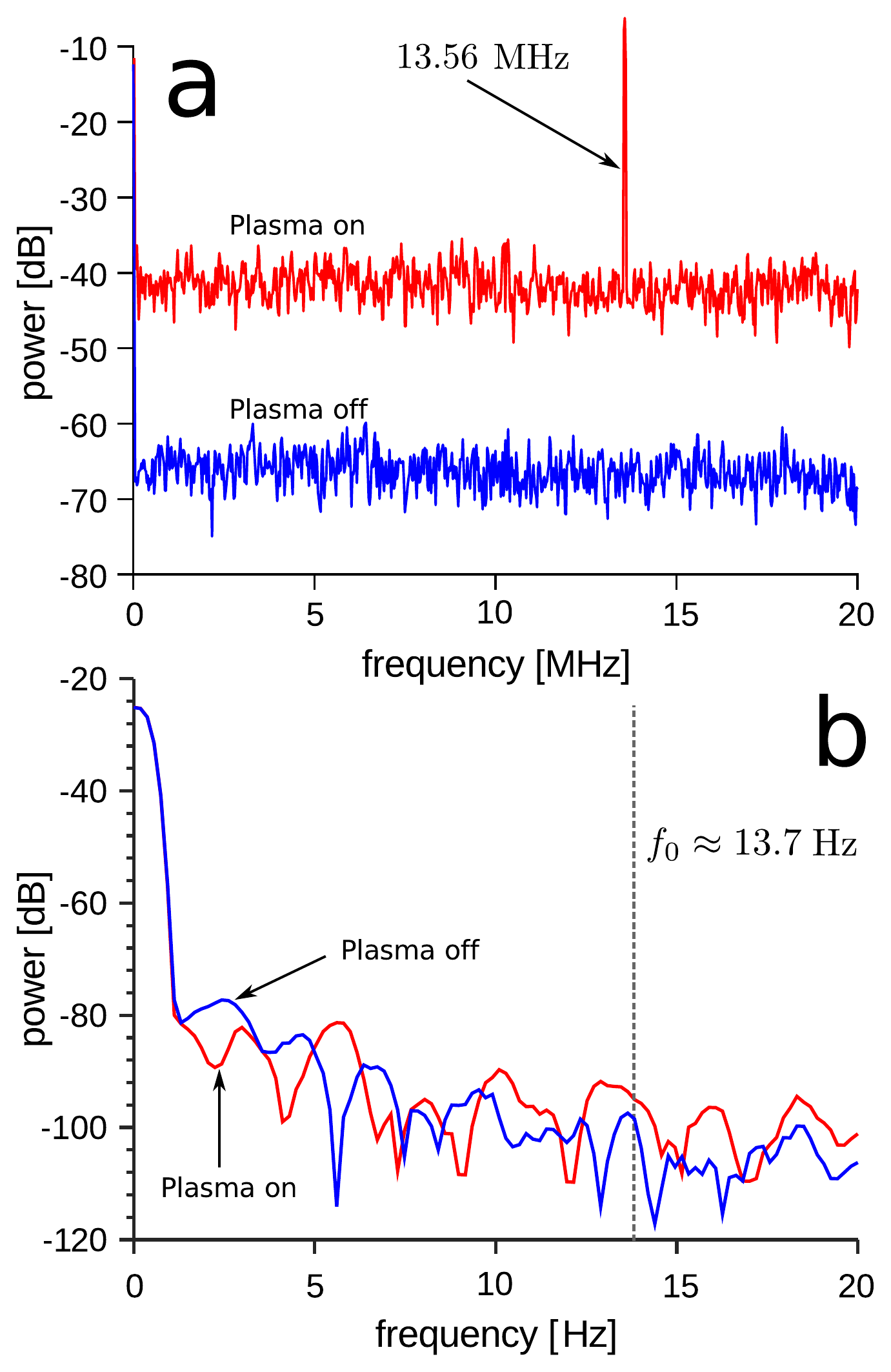}
	\caption{\label{fourier_bias} (a) Broadband spectrogram of the bias voltage $\phi_\text{dc}$ on the electrode with the plasma on (red; note the power spike at 13.56 MHz) and off (blue). The pressure was $P$ = 0.6 Pa. Note that the ``off'' has been shifted downward for clarity. (b) Narrow band spectrogram the bias voltage $\phi_\text{dc}$ on the electrode with the plasma on (red) and off (blue) at frequencies near characteristic frequencies of particle oscillation (Sec.\ \ref{charge_mod}).}
	
\end{figure} 

Beyond extracting fundamental plasma parameters, we employed our Langmuir probe to determine the existence of variations in the plasma under the conditions where particle oscillations are observable. For a fixed $\phi_\text{b}$, the probe can resolve very small changes in the plasma environment. For instance, Fig.\ \ref{fourier_plasma}a shows the variation of the probe current ($\phi_\text{b}$ = 15 V $\pm$ 100 mV, 40 mm above the powered electrode) as $\phi_\text{dc}$ is varied a small amount (as measured by a high-resolution voltmeter). The homogeneity of our plasma at low frequencies can be quantitatively demonstrated by the Fourier transform of the Langmuir probe current in Fig.\ \ref{fourier_plasma}b. Here, the probe was placed 15 mm above the electrode (near the edge of the sheath) and biased at $\phi_\text{b}$ = 0 V. Between 0.1 and 1000 Hz, there are no discernible spectral components that would give rise to particle oscillations. In addition to searching for variations in the probe current, we obtained spectrograms (using an HP spectrum analyzer) of the the bias voltage $\phi_\text{dc}$ to determine whether fluctuations on the electrode were present at conditions under which particles oscillate. The broadband spectrograms of $\phi_\text{dc}$ illustrated in Fig.\ \ref{fourier_bias}a show variations in voltage on the powered electrode with the plasma on (red curve) and the plasma off (blue curve). Note the power spike at 13.56 MHz for the ``on'' curve. Both curves have the same base level, but the ``on'' curve has been shifted upward for clarity. Similarly, Fig.\ \ref{fourier_bias}b shows spectrograms of $\phi_\text{dc}$ with the plasma on (red curve) and off (blue curve), but at frequencies close to the frequency of particle oscillations. Both ``on'' and ``off'' narrow band curves are quantitatively similar and do not show features that would account for particle oscillations at frequencies below 20 Hz (Fig.\ \ref{mean_peak}). We will return to the subject of plasma fluctuations in Sec.\ \ref{sto_flu} in greater detail. However, the measurements addressed presently suggest that it is unlikely that temporal inhomogeneities in the plasma are responsible for the spontaneous vertical oscillations observed in our experiments. 

\section{Analysis and Discussion}
\subsection{Forces on a dust particle}
\label{forces}
In order to understand the origin of the spontaneous oscillations, it is necessary to enumerate the forces acting on a particle in the vertical direction \cite{Melzer2008,Nitter1994}. First, gravity acts on the particle with force $F_\text{g}=-m_\text{p} g$, where $m_\text{p}$ is the mass of the particle and $g$ is the acceleration due to gravity. For the 9.46 $\mu$m MF particles used in our experiment with density $\rho_\text{p}$ = 1510 kg/m$^3$, $F_\text{g}\approx -6.6 \times 10^{-12}$ N. The electrostatic force acting against gravity is given by 
\begin{equation}
F_\text{e}=Q_\text{eq}E.
\end{equation}
Here $E$ is the electric field in the sheath, which points in the negative $z$ direction, and depends on position $z$. $Q_\text{eq}$ refers to the average charge the particle will obtain at a given position $z$ in the sheath given sufficient time. This will later become important when the particle moves rapidly through the sheath (Sec.\ \ref{delayed}). 

The particle will also feel a drag force as it moves through the background of the neutral gas:
\begin{equation}
F_\text{d}=-\gamma m_\text{p} v_\text{p},
\label{Fd}
\end{equation}
where $v_\text{p}$ is the velocity of the particle in the $z$-direction. The damping rate $\gamma$ is given by the Epstein law \cite{Melzer2008}:
\begin{equation}
\gamma=\delta\dfrac{2P}{a_\text{p}\rho_\text{p}}\sqrt{\dfrac{2 m_\text{n}}{\pi k T_\text{n}}},
\label{damp}
\end{equation}
where $m_\text{n}$ and $T_\text{n}$ are the mass and temperature of the neutral gas species, respectively. In our experiments with argon, $m_\text{n}=6.64 \times 10^{-26}$ kg and $T_\text{n}$ = 298 K. The coefficient $\delta$ ranges from 1.0-1.44 depending on the nature of scattering of neutral atoms, namely specular vs. diffuse reflection. Recent experiments have shown that $\delta\approx 1.44\pm0.05$ \cite{Jung2015}, so we used $\delta=1.44$ for all calculations. For MF particles moving at $P$ = 0.6 Pa, $\gamma$ = 0.79 s$^{-1}$. 

We also consider the ion drag force, $F_\text{i}$, from the rapidly moving ions that are accelerated towards the electrode  \cite{Melzer2008,Zafiu2003,Ivlev2004,Semenov2013}. The ion drag force may be relevant for our experiments since it contains a non-monotonic dependence on the relative dust-ion velocity, and potentially, spontaneous oscillations of dust particles through ``negative damping'' in the regime where the force \textit{decreases} with increasing ion velocity \cite{Samarian2001,Nitter1996}. The ion drag force has two components, a collection force, $F_\text{dir}$, due to direct collisions, and a Coulomb force, $F_\text{Coul}$, from ions scattered from the Debye shield around the particle \cite{Melzer2019b,Khrapak2005,Patacchini2008,Barnes1992,Ivlev2004,Hutchinson2006}. In the simple Barne's model of a collisionless plasma \cite{Barnes1992,Melzer2019b}, these forces are
\begin{eqnarray}
\label{iondrag1}
F_\text{dir}&=-\pi b_\text{c}^2 m_\text{i}n_\text{i}v_\text{i}v_s,\\
F_\text{Coul}&=-2\pi b_{\pi/2}^2 m_\text{i}n_\text{i}v_\text{i}v_s\ln\left(\dfrac{\lambda_\text{D}^2+b_{\pi/2}^2}{b_\text{c}^2+b_{\pi/2}^2}\right),
\label{iondrag2}
\end{eqnarray}
where $v_\text{s}^2=v_\text{i}^2+v_\text{th,i}^2$, $v_\text{i}$ is the ion drift speed, $v_\text{th,i}$ is the ion thermal speed, $b_{\pi/2}=Q_\text{eq}e/4\pi\epsilon_0m_\text{i}v_\text{s}^2$ is the impact parameter for perpendicular scattering,  $a_\text{p}$ is the radius of the particle, and $b_\text{c}=a_\text{p}(1-2b_{\pi/2}/a_\text{p})^{1/2}$ is the minimum collision parameter. Often the electron Debye length (Eq.\ \ref{eq4.1}) is used for $\lambda_\text{D}$ as a rough approximation. 

A more sophisticated model \cite{Hutchinson2006,Khrapak2005,Khrapak2002,Melzer2019b} builds on this approach by considering scattering outside the Debye sphere and shifted Maxwellian ion velocity distributions. The result is a decrease in the ion drag force by a factor of $\approx$ 2. However, assuming the Barnes et al. \cite{Barnes1992} model above (Eqs.\ \ref{iondrag1} and \ref{iondrag2}), we estimate the total ion drag force to be approximately one order of magnitude smaller than the gravitational and electrostatic force. For example, as an upper bound on this force in our argon environment, we assume that $Q_\text{eq}\approx -35,000e$ (see Sec.\ \ref{charge_mod}), $v_\text{i}\approx 2000$ m/s (Bohm velocity), $v_\text{th,i}\approx 400$ m/s, $n_\text{i}\approx5\times10^{13}$ m$^{-3}$, and the logarithm in Eq.\ \ref{iondrag2} is $\approx 4$, the total ion drag force is 9\% of the gravitational force. In many experiments, the ion density is smaller and the ion velocity is larger than the Bohm velocity in the sheath. 

Additionally, if we consider the regime where the ion drag force decreases with the ion velocity (where ``negative damping'' is possible \cite{Nitter1994}), the size of this effect is 3 orders of magnitude too small to explain the oscillations we observe. Physically, if we consider a particle moving downward toward the electrode, in the same direction as the ions, then its relative velocity with respect to the ions will be slightly smaller, and the ion drag force will be slightly larger, pushing it down with a larger force. This can be seen quantitatively by replacing $v_\text{i}$ with $v_\text{i}+v_\text{p}$, where $v_\text{p}$ is the particle velocity, and linearizing around $v_\text{p}=0$. The result is a force $-m_\text{p}\gamma_\text{n}v_\text{p}$, where $\gamma_\text{n}$ is an effective ``negative'' damping coefficient of order $-10^{-3}$ s$^{-1}$, which is much smaller in magnitude than the neutral gas damping coefficient in our experiments: $\gamma\approx$ 0.8-1.6 s$^{-1}$.

%

\begin{figure*}[t]
	\centering
	\includegraphics[width=6.5in]{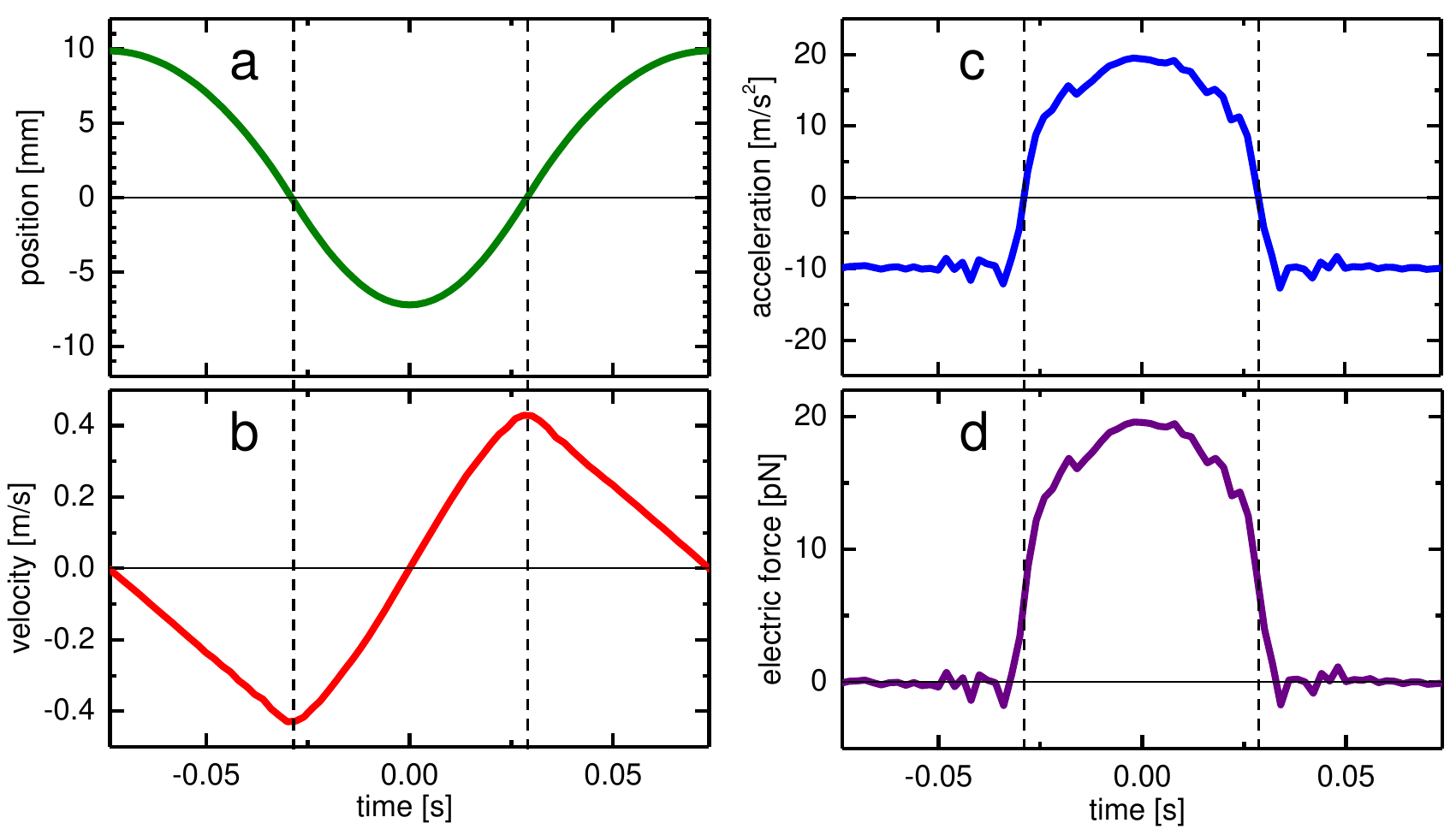}
	\caption{\label{motion_analysis} (a) Vertical position vs. time for a 9.46 $\mu$m diameter MF particle at $P$ = 0.61 Pa. Data has been averaged over 400 oscillation cycles. The equilibrium position of the particle was 14 mm above the electrode surface, and $\phi_{\text{w}}$ = -28 V ($\phi_\text{dc}$ = -6 V). The velocity (b) and acceleration (c) are computed from derivatives of the data. (d) Electrostatic force computed from the data, assuming Eq.\ \ref{zeq}. The dashed lines represent zero net acceleration, where gravity and electrostatic forces are approximately balanced.}
\end{figure*} 

\subsection{Estimate of particle charge}
\label{charge_mod}

Although Langmuir probe measurements have been used to estimate the floating potential deep in the plasma sheath \cite{Faudot2019}, these measurements may not be reliable enough to estimate the particle charge in our experiments since the probe itself disturbs the sheath environment. Thus, we chose to extract the charge from dynamical measurements of the particles' motion under large-amplitude oscillations. Figure \ref{motion_analysis}a shows one cycle of the $z$-position vs. time for a single particle oscillating at $P$ = 0.61 Pa. The data for each cycle was linearly interpolated at the same time points and averaged over 400 cycles of the steady oscillation in order to reduce noise in calculating derivatives. The velocity and acceleration associated with the position are shown in Fig.\ \ref{motion_analysis}b-c. The position has been adjusted so that zero corresponds to the maximum velocity where the net restoring force is zero. The error in the acceleration is largest at the sheath edge ($t\approx\pm0.03$ s), which is due to the slope discontinuity in the velocity. This results in small oscillations visible in the data. 

First, we note that for such large-amplitude oscillations, the particle spends a significant amount of time above the sheath, where the net acceleration is due to gravity, $\approx$ -9.8 m/s$^2$. Due to the damping effects of the neutral gas (Eq.\ \ref{damp}), the particle must receive a net ``kick" within the sheath in order to maintain a nearly constant amplitude of oscillation over minutes. If gravity, hydrodynamic damping, and electrostatic forces are the only forces acting on the particle, then the electrostatic force must not be purely conservative, i.e. the charge is not solely a function of position. The time dependence of $Q$ will be discussed in more detail in Sec.\ \ref{delayed}.

In order to estimate the particle charge, first we extract the net electrostatic force assuming the equation of motion for the $z$-position is:
\begin{equation}
m_\text{p}\ddot{z}=-m_\text{p} \gamma \dot{z}-m_\text{p} g+E(z)Q(z,t),
\label{zeq}
\end{equation}
where $E(z)$ is the spatially-varying electric field in the vertical direction and $Q(z,t)$ is the charge on the particle, which varies with both $z$ and $t$. Using the velocity and acceleration shown in Fig.\ \ref{motion_analysis}b-c, we compute $E(z)Q(z,t)$ as a function of time. The result is shown in Fig.\ \ref{motion_analysis}d. To a good approximation, the force is symmetric about $t=0$, so to a good approximation, we can assume that $Q=Q(z)$ for now.  

\begin{figure}[t]
	\centering
	\includegraphics[width=\columnwidth]{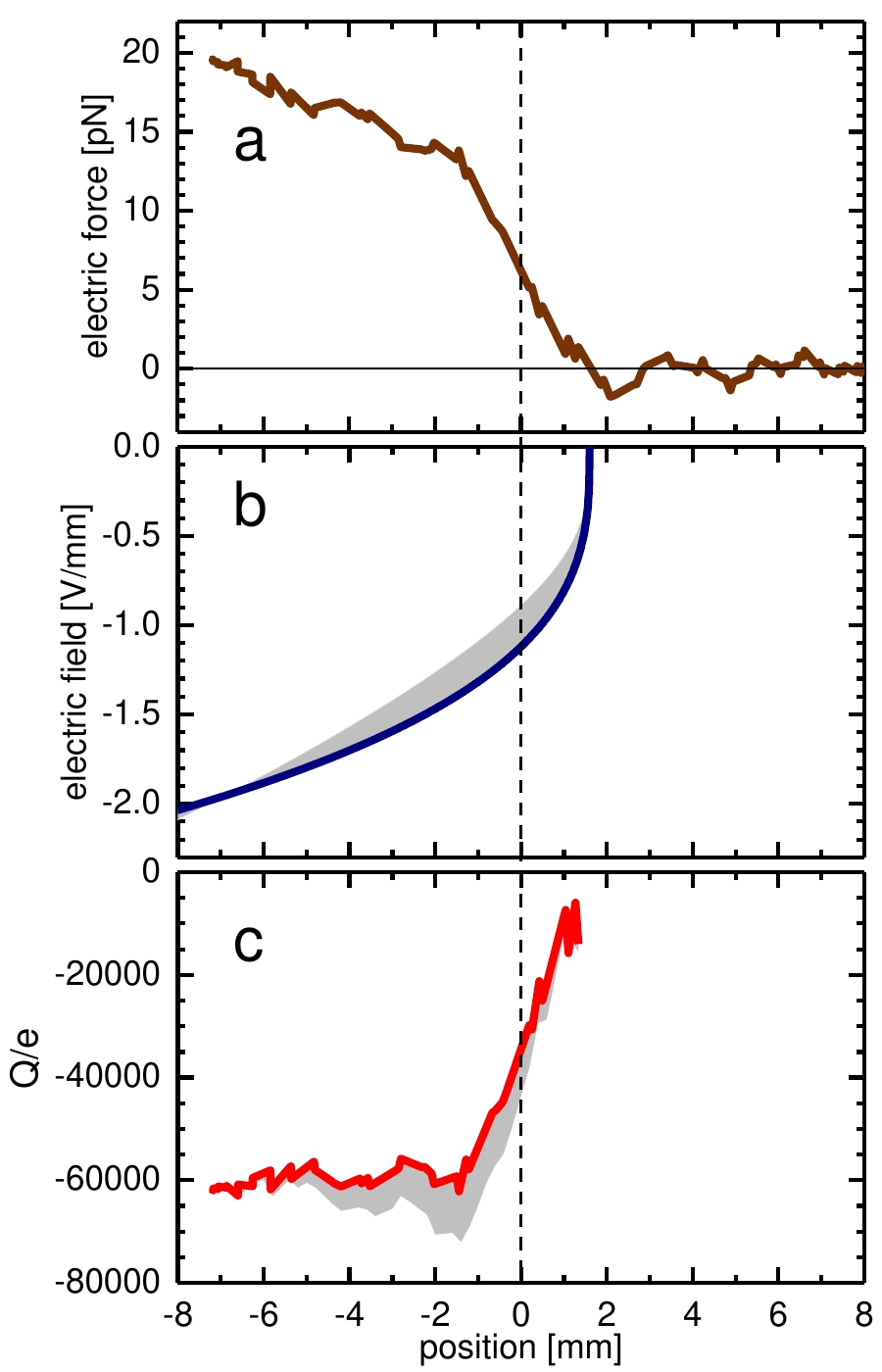}
	\caption{\label{efield_charge} (a) Electrostatic force vs. position for an 9.46 $\mu$m MF particle at $P$ = 0.61 Pa and $\phi_\text{w}$ = -28 V. The equilibrium position corresponds to $z=0$. The edge of the sheath occurs at $z\approx$ 1.6 $\pm$ 0.1 mm. The oscillations in the data near sheath's edge are due to differentiation of the numerical position data. (b) Electric field computed from Eq.\ \ref{child_lang} with $\phi_\text{w}$ = -28 V, $z_\text{s}$ = 1.6 mm, and $z_\text{w}$ = -14 mm. (c) Equilibrium charge, normalized by the elementary charge $e$. This was computed by dividing the data in (a) by the electric field in (b).}
\end{figure}

The total electric force is plotted vs. position in Fig.\ \ref{efield_charge}a. 
At the equilibrium position, as shown by the vertical dashed line, we can compute the angular frequency of small oscillations, $\omega_0$:
\begin{equation}
\omega_0^2m_\text{p}= -\dfrac{d\left(E(z)Q_\text{eq}(z)\right)}{dz}\bigg\rvert_{z=0},
\label{freq_0}
\end{equation}
where $\omega_0=2\pi f_0$. For the data shown in Fig.\ \ref{efield_charge}a, $f_0\approx$ 13.7 Hz. The actual frequency of the anharmonic particle motion is significantly smaller than this because of the amount of time the particle spends in free fall above the edge of the sheath. The edge of the sheath is determined by the points where total electric force goes to zero, which is approximately 1.6 $\pm$ 0.1 mm above the equilibrium position ($z=0$ mm).

In order to proceed further, we need a model for the electric field, $E(z)$. At higher pressures, self-consistent fluid models for the electrons and ions in the sheath are more accurate for determining the electric field \cite{Wang1999,Land2010,Douglass2011,Douglass2012,Sheridan1991}, but rely on more parameters which are difficult to measure experimentally. At lower pressures and larger voltages, the sheath is ion-dominated. Thus, we used perhaps the simplest analytical model for the sheath, which is the Child-Langmuir law \cite{Benilov2009}:
\begin{equation}
\label{child_lang}
E(z)=\dfrac{4\phi_\text{w}(1-z/z_\text{s})^{1/3}}{3z_\text{s}(1-z_\text{w}/z_\text{s})^{4/3}},
\end{equation}
where $\phi_{\text{w}}=\phi_{\text{dc}}-\phi_{\text{p}}$ is the potential on the conducting wall relative to the plasma potential, $z_\text{w}$ is the position of the wall, and $z_\text{s}$ is the position of the sheath boundary. Technically, this model is valid for low-pressure, collisionless plasmas where 1) the electrons are significantly depleted in the sheath, and when 2) the mean free path of the ions ($\lambda_\text{in}$, Eq.\ \ref{inmf}) is much larger than the spatial extent of the sheath \cite{Miller1997}. Since the total potential difference across the electrode sheath is rather large, $e\left|\phi_\text{w}\right|/k_\text{B}T_\text{e}\approx 30$, the first condition is reasonably satisfied for most of the sheath. 

However, for the low pressures in our experiments ($0.6<P<1.3$ Pa), the mean free path ranges from 6.7 mm $<\lambda_\text{in}<$ 3.1 mm. This is comparable to the extent of the particles' motion in the sheath. Thus, in order to estimate the effects of collisions on the electric field (i.e. deviations from Eq.\ \ref{child_lang}), we used the ion-dominated model described in Naggary et al. \cite{Naggary2019}. Poisson's equation determines the electric potential, and the one-dimensional ion motion is described by the continuity equation. Ions are also accelerated by the electric field, and lose momentum through collisions:
\begin{eqnarray}
\label{collis1}
\dfrac{dE}{dz}=&\dfrac{e\Psi}{\epsilon_0 v_\text{i}},\\
m_\text{i}v_\text{i}\dfrac{dv_\text{i}}{dz}=&eE-\dfrac{|v_\text{i}|}{\lambda_\text{in}}m_\text{i}v_\text{i}.
\label{collis2}
\end{eqnarray}
Here, $\Psi$ is the flux of ions and $v_\text{i}$ is the velocity. Both quantities are negative since ion flow in the negative $z$-direction towards the electrode. To generate electric field profiles, we solved Eqs.\ \ref{collis1} and \ref{collis2} numerically with the boundary conditions $E(z_\text{s})=0$, $v_\text{i}(z_\text{s})=0$. The value of $\Psi$ was adjusted so that the total potential difference across the sheath was $\phi_\text{w}$:
\begin{equation}
\int_{z_\text{w}}^{z_\text{s}}Edz=\phi_\text{w}.
\end{equation}

For the data shown in Fig.\ \ref{motion_analysis}, $\phi_\text{w}$ = -28 V, $z_\text{w}$ = -14 mm, and $z_\text{s}\approx$ 1.6 mm. With these parameters, the electric field is shown in Fig.\ \ref{efield_charge}b. The solid line is the solution given by Eq. \ref{child_lang}, and the shaded region represents solutions of Eqs. \ref{collis1} and \ref{collis2} where 2 mm $<\lambda_\text{in}<\infty$. By dividing the electrostatic force by the electric field, we arrive at an estimate of the particle charge. Figure \ref{efield_charge}c shows the $Q/e$ as a function of position. Again, the shaded region represents the collisional solution. The charge is relatively constant in the middle of the sheath, near 60,000--70,000 electrons, and decreases to $\approx$ 10,000 electrons at the sheath's edge. Although we expect the equilibrium number of electrons to decrease further into the sheath (below $z$ = -8 mm) since the electrons will be significantly depleted, the relatively flat region of constant charge agrees with previous measurements of dust charge in plasma sheaths \cite{Douglass2012}. The positive slope of the charge near the equilibrium position, $dQ/dz>0$, is an important feature that will be discussed in Sec.\ \ref{delayed}. Although collisions do have a non-negligible effect, for pedagogical simplicity, we will proceed with the analytical Child-Langmuir law in the remaining discussion.

At the particle's equilibrium position ($z=0$), the charge is approximately 35,000 electrons. This agrees extremely well with orbital-motion-limited (OML) theory predictions for $\approx$ 10 $\mu$m particles, albeit at higher pressures (20 Pa) \cite{Douglass2011,Douglass2012}. At the low pressures used in our experiments, it is possible that electron emission from the electrode enhances the electron temperature in the sheath, which would lead to a larger negative charge than expected using OML theory \cite{Ingram1988,Mussenbrock2006}. In Douglass et al. \cite{Douglass2011}, OML theory is applied in the sheath using a shifted Maxwellian distribution of ion velocities and a fluid model for both the electron and ion densities in the sheath. These models usually produce a smoother, extended transition from the sheath to the pre-sheath and bulk plasma. Here we have used the Child-Langmuir law in order to estimate the particle charge, which has a very sharp, concave transition (Fig.\ \ref{efield_charge}b). This sharp transition is consistent with rf sheath measurements at lower pressures, for example, in Krypton at $P=1$ Pa \cite{Czarnetzki2013}. The validity of this model may be poor at the sheath's edge where the electron's are not yet depleted. However, given that the force drops nearly linearly to zero (Fig.\ \ref{efield_charge}a) where the particle is essentially in free fall (Fig.\ \ref{motion_analysis}c), we surmise that the Child-Langmuir law is reasonable for the plasma environment in our experiments.

\subsection{Stochastic fluctuations}
\label{sto_flu}
A number of authors have proposed stochastic fluctuations, either of the plasma environment or of the charge on the particle itself, as a mechanism to generate oscillations \cite{Sorasio2002b,Ivlev2000Inf}. In either case, such fluctuations could turn the particle into a randomly-forced harmonic oscillator. However, this is inconsistent with our observations since stochastic forcing invariably gives rise to large variations in the amplitude of oscillation. To show this, we used stochastic numerical simulations to investigate the effect of plasma environment and particle charge fluctuations on the vertical oscillations. First, we characterized the effect of sheath boundary fluctuations on the vertical oscillations. We assume the particle motion obeys the following equation:
\begin{equation}
m_\text{p}\ddot{z}=-m_\text{p} \gamma \dot{z}-m_\text{p} g+E(z,t)Q(t),
\label{stoch1}
\end{equation} 
where $E(z,t)$ is an electric field given by the Child-Langmuir law (Eq.\ref{child_lang}), and $Q(t)$ is the particle charge. The particle experiences stochastic forcing over time and the relative size of these fluctuations depends on the sheath boundary location, $z_{s}(t)$, which is a normally-distributed, uncorrelated random variable with mean $\langle z_\text{s}\rangle$ and variance $\delta z_\text{s}^2$. In this case, $Q(t)$ is constant in time and equal to $\approx$ $-36,000e$ in the simulation.


\begin{figure}[!] 
	\centering
	\includegraphics[width=\columnwidth]{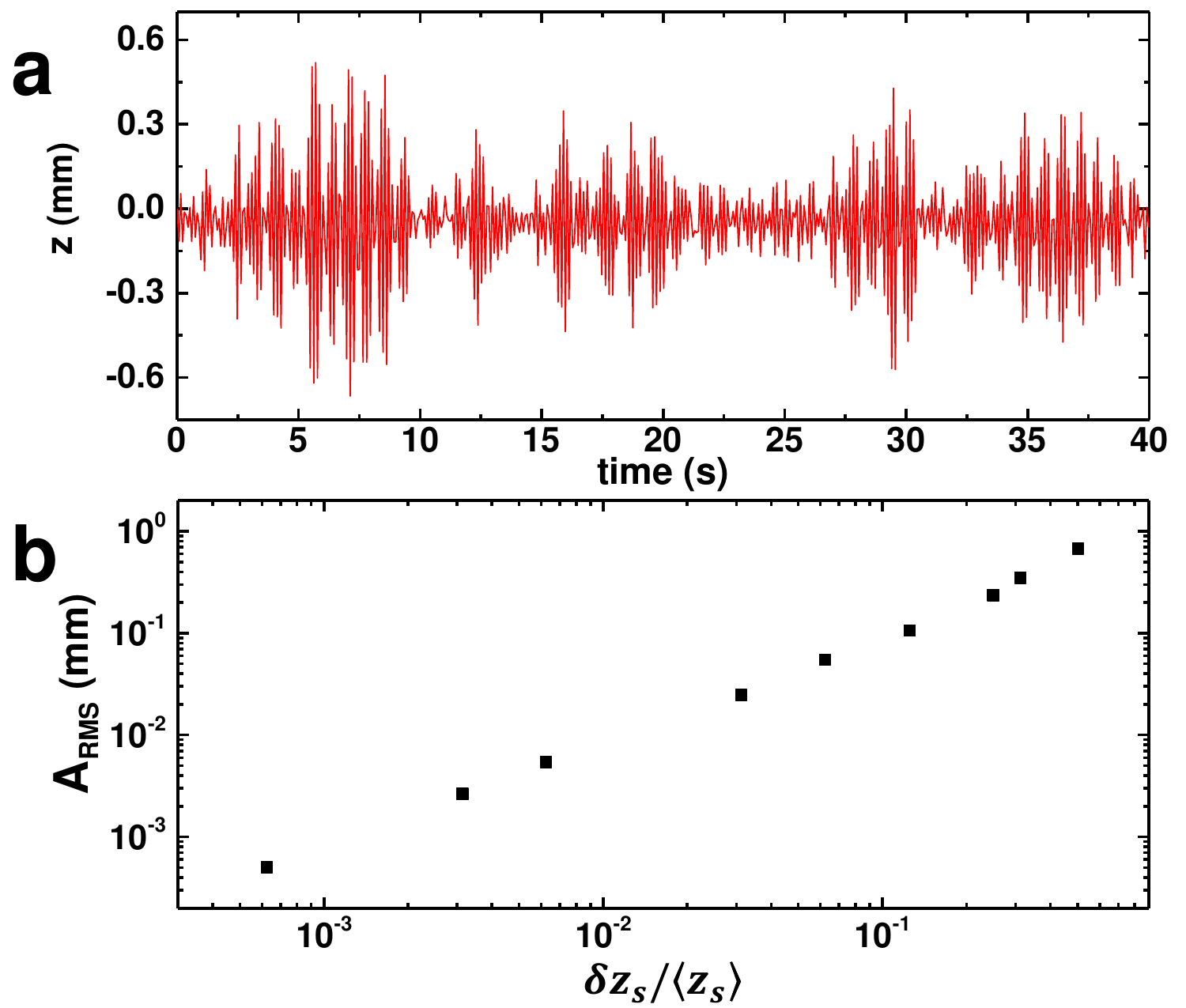}
	\caption{\label{stochastic_zs}(a) The trajectory of a single particle driven by stochastic variations of the sheath boundary ($z_\text{s}$ in Eq.\ \ref{child_lang}). Parameters for the Child-Langmuir law were $\phi_\text{w}$ = -28 V, $z_\text{w}$ = -14 mm, and $z_\text{s}=$ 1.6 mm. The damping constant $\gamma$ was set to 0.68 s$^{-1}$ and $m_\text{p}=7.85\times10^{-13}$ kg. (b) Root mean squared (rms) amplitude of vertical oscillations as a function of the normalized sheath boundary fluctuations.}
\end{figure} 

 Stochastic fluctuations of the sheath boundary resulted in spontaneous oscillations of a single particle (Fig.\ \ref{stochastic_zs}a). Notably, the amplitude varies over time, as expected for a stochastically-driven oscillator, drawing a striking difference from our experimental observations (Fig.\ \ref{fourier}a). Fig. \ref{stochastic_zs}b shows the rms (root mean squared) amplitude of the oscillations as a function of the normalized sheath boundary fluctuations, $\delta z_\text{s}/\langle z_\text{s}\rangle$. Importantly, an oscillation with a conservative amplitude of 1 mm requires stochastic sheath variations of nearly 60\%, meaning that the sheath boundary would fluctuate by $\approx$ 1 mm. Although this is plausible, these fluctuations would be visible withing the plasma environment, and we did not observed any visible changes below the threshold pressure needed to induce oscillations. 
 
 \begin{figure}[!] 
	\centering
	\includegraphics[width=\columnwidth]{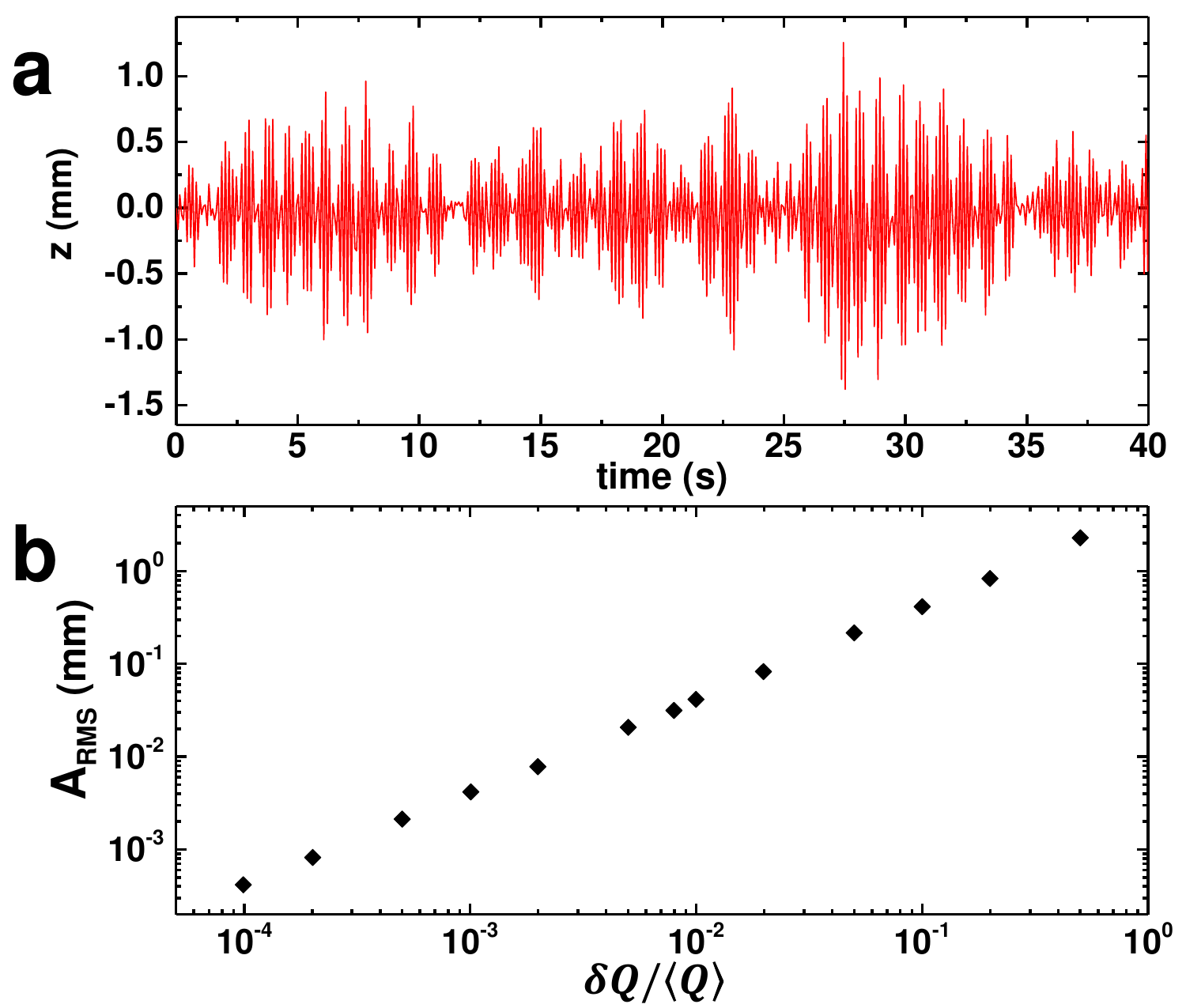}
	\caption{\label{stochastic_charge}   (a) The trajectory of a single particle driven by stochastic fluctuations of the particle charge. The parameters for the Child-Langmuir law (Eq.\ \ref{child_lang}) were $\phi_\text{w}$ = -28 V, $z_\text{w}$ = -14 mm, and $z_\text{s}=$ 1.6 mm. The damping constant $\gamma$ was set to 0.68 s$^{-1}$ and $m_\text{p}=7.85\times10^{-13}$ kg.  (b) Root mean squared (rms) amplitude of vertical oscillations as a function of the charge fluctuations.}
\end{figure} 
 
 Furthermore, we investigated the effect of charge fluctuations on the vertical oscillations, assuming the sheath position was fixed. We utilized Eq.\ \ref{stoch1} with one modification: the charge was not constant but rather experienced temporal fluctuations. The fluctuations occur around mean $\langle Q \rangle$ with  variance of  $\delta Q^2$. Fig.\ \ref{stochastic_charge}a shows the temporal evolution of the particle for $\delta Q/\langle Q \rangle \approx 0.1.$ As Fig.\ \ref{stochastic_charge}b illustrates, even for relative charge fluctuations of order 100\%, the rms amplitude of oscillations is an order of magnitude smaller than the experimentally observed values. For either mechanism, the required fluctuations are too large and the amplitude too variable to be consistent with our experiments. 
 
\subsection{Delayed charging}
\label{delayed}

The experimental data points to a potential mechanism for the large amplitude oscillations without stochastic processes. As first introduced by Nunomura et al. \cite{Nunomura1999}, a positive gradient in the equilibrium charge profile can couple to the finite charging time of the particle and produce a net positive work during a single oscillation cycle. However, the amplitude of the oscillations reported in Nunomura et al. \cite{Nunomura1999} were less than 1 mm. A linear stability analysis later showed that the condition for the inception of the instability is an  effective negative damping constant \cite{Ivlev2000Inf}:
\begin{equation}
\gamma_\text{eff}=\gamma-\dfrac{1}{2}\left(\dfrac{EQ_\text{eq}'}{(Q_\text{eq}E)'}\right)\dfrac{\omega_0^2}{\nu},
\end{equation}
where the primes denote differentiation with respect to $z$ and all quantities are evaluated at the equilibrium position. Here, the fundamental angular frequency of small oscillations is $\omega_0=2\pi f_0$, and $\nu$ is the charging frequency determined by the collection of ions and electrons on the particle. To linear order, when $\gamma_\text{eff}$ is negative, the oscillations will increase exponentially without bound. Obviously, the particle oscillations reach some maximum amplitude due to changes in the equilibrium charge profile (i.e. $Q_\text{eq}(z)$). As shown in Fig.\ \ref{motion_analysis}d, the particles can exit the sheath entirely for much of their oscillation cycle, where the electric force is zero. Here we combine a simple model for the potential in the sheath and the equilibrium charge with the charging dynamics model from Ivlev et al. \cite{Ivlev2000Inf} to directly compare with our the experimental measurements.

Specifically, we model the vertical motion of a single, charged particle in a 1D spatially-varying electric field with a time-dependent charge \cite{Ivlev2000NL,Ivlev2000Inf}. As in Eq.\ \ref{zeq}, the equation of motion for the particle's vertical position is:
\begin{equation}
m_\text{p}\ddot{z}=-m_\text{p} \gamma \dot{z}-m_\text{p} g+E(z)Q(z,t),
\label{zeq2}
\end{equation}
where $E(z)$ is the spatially-varying electric field in the vertical direction and $Q(z,t)$ is the time-varying charge on the particle, which intrinsically depends on the particle position. The vertical position of the particle is $z(t)$, where $z=0$ is the equilibrium position determined by electrostatic and gravitational forces. 

The charge on the particle follows a simple exponential decay towards its equilibrium value:
\begin{equation}
\dot{Q}=-\nu (Q-Q_\text{eq}(z)),
\label{qeq}
\end{equation}
In the limit $\nu/\omega_0\rightarrow\infty$, the particle always remains at the position-dependent equilibrium charge, $Q_\text{eq}(z)$, and a closed path of motion can only result in zero net work done on the particle. However, even if $\nu/\omega_0\approx 100$, this is still sufficient to cause large-amplitude oscillations, provided that $Q_\text{eq}'(0)>0$ and is sufficiently large.

In order to quantitatively interpret the experimental data for a single particle oscillation, we require a spatial model for the electric field in the sheath, $E(z)$, and the equilibrium charge on the particle, $Q_{eq}(z)$. As before (Eq.\ \ref{child_lang}), we assume the simplest model for $E(z)$ by using the Child-Langmuir law:
\begin{equation}
E(z)=\dfrac{4\phi_\text{w}(1-z/z_\text{s})^{1/3}}{3z_\text{s}(1-z_\text{w}/z_\text{s})^{4/3}},
\label{ch_lang}
\end{equation}
where $\phi_{\text{w}}=\phi_{\text{dc}}-\phi_{\text{p}}$ is the potential on the conducting wall relative to the plasma potential, $z_\text{w}$ is the position of the wall, and $z_\text{s}$ is the position of the sheath boundary. Based on our measurements of the particle charge from Fig.\ \ref{efield_charge}b, we choose a simple cubic function for the equilibrium charge for $-z_\text{s}<z<z_\text{s}$:
\begin{eqnarray}
Q_\text{eq}(z)=&Q_0+Q_1z-\dfrac{Q_1z^3}{3 z_\text{s}^2}
\end{eqnarray}
%
This form ensures a \textit{positive} slope for the charge at $z=0$ and zero slope at $z=\pm z_\text{s}$. We assume that the equilibrium charge is constant above and below these values, i.e. $Q_\text{eq}(z)=Q_\text{eq}(z_\text{s})$ for $z>z_\text{s}$ and $Q_\text{eq}(z)=Q_\text{eq}(-z_\text{s})$ for $z<-z_\text{s}$. The values of $Q_0$ and $Q_1$ are constrained by the two equilibrium conditions
\begin{eqnarray}
\omega_0^2m_\text{p}=&-\dfrac{d\left(E(z)Q_\text{eq}(z)\right)}{dz}\bigg\rvert_{z=0},\\
m_\text{p} g=&E(0)Q_\text{eq}(0).\nonumber
\end{eqnarray}
Using these conditions, and solving for $Q_0$ and $Q_1$ results in
\begin{eqnarray}
Q_0=&\dfrac{3 m_\text{p} g(z_\text{s}-z_\text{w})^{4/3}}{4\phi_\text{w}z_\text{s}^{1/3}},\\
Q_1=&-\dfrac{m_\text{p} (z_\text{s}-z_\text{w})^{4/3}(3z_\text{s}\omega_0^2-g)}{4\phi_\text{w}z_\text{s}^{4/3}}.\nonumber
\end{eqnarray}
The parameters $\phi_\text{w}$, $z_\text{w}$, $z_\text{s}$, and $\omega_0$ are all tightly constrained by experimental measurement (i.e. Figs.\ \ref{motion_analysis} and \ref{efield_charge}). Thus, in order to fit the data, only $\nu$ is a truly adjustable parameter. 

\begin{figure}[!]
	\centering
	\includegraphics[width=\columnwidth]{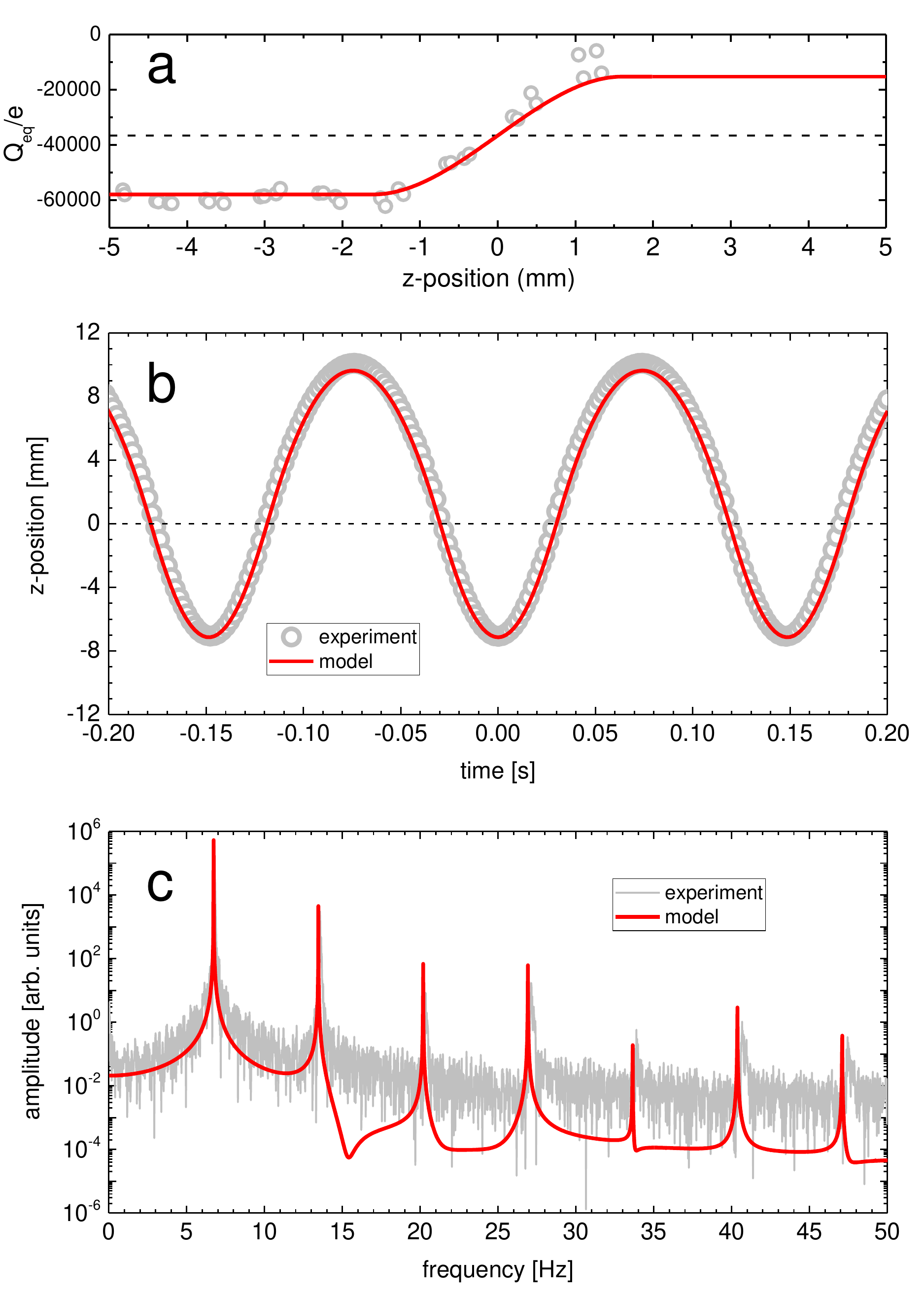}
	\caption{\label{model_fit} (a) Equilibrium charge (normalized by $e$) as a function of $z$ for the delayed charging model. The equilibrium position of the particle is $z=0$, and the edge of the sheath is approximately $z=1.6$ mm. The gray data points are the measurements from Fig.\ \ref{efield_charge}c. (b) Vertical position of a single MF particle (9.46 $\mu$m diameter) as a function of time (gray points). The pressure in the chamber was $P$ = 0.6 Pa, and the bias was $\phi_{\text{dc}}$ = $-6$ V. The red line is the best fit using the delayed charging model. (c) Fourier transforms of the experimental data (similar to Fig.\ \ref{fourier}b, gray line) and the model output (solid red line).}
\end{figure}

Figure \ref{model_fit} shows results from the nonlinear regression using time series generated by the model with an MF particle of diameter 9.46 $\mu$m, $\phi_\text{w}$ = $-28$ V ($\phi_{\text{dc}}=-6$ V), $\gamma$ = 0.79 s$^{-1}$, $f_0=\omega_0/2\pi=13.7$ Hz, $z_\text{w}=-14$ mm, and $z_\text{s}$ = 1.6 mm. The spatial variation of the equilibrium charge agrees well with our estimates of the particle charge from the analysis of the particle motion shown in Fig.\ \ref{efield_charge}c. This provides confidence in the model, and allows for a quantitative estimate of the particle charge, the charging time, and the spatial variation of the equilibrium charge. 

Moreover, the model provides an excellent fit to large amplitude oscillation data, as shown in Fig.\ \ref{model_fit}b. We fit the data by solving Eqs.\ \ref{zeq} and \ref{qeq} with initial conditions $z(0)$ = 7 mm, $\dot{z}(0)$ = 0 mm/s, and $Q(0)$ = -30,000$e$ using built-in routines in \textit{Mathematica}, then fit the solution with a few cycles of the experimental data once the solution has reached a steady-state value (usually after $t\approx$ 30 s). The arbitrary phase offset between the solution and the experiment is automatically adjusted to maximize the quality of the fit. We obtain similar fits for all of our oscillation data. 

We again note here that the plasma sheath ends at $z=z_\text{s}\approx$ 1.6 mm (Fig.\ \ref{model_fit}a). However, the maximum vertical position is larger that 10 mm. This means that for much of the cycle of motion, the particle is essentially in free fall, with the addition of drag from the neutral gas. As the particle descends further into the sheath, both the magnitude of the electric field and the charge increase, so the total upward force increases significantly, leading to a sort of ``bouncing'' effect. This explains the strong asymmetry of the motion, and the presence of multiple harmonics in Fig.\ \ref{model_fit}b. The model is able to quantitatively capture the entire shape of the Fourier spectrum with no adjustable parameters. Our nonlinear regression procedure uses the sum of the squares of the residuals ($\chi^2$) as a quality of the fit. The 95\% confidence interval for the best fit in Fig.\ \ref{model_fit} is $\nu$ = 1133 $\pm$ 50 s$^{-1}$. Thus the characteristic charging time for the particle is $\nu^{-1}\approx$ 880 $\mu$s, which is quite large, but not unrealistic given the environmental conditions (low pressure, low ion and electron density, and relatively low $T_\text{e}$), and is in excellent agreement with estimates of the slower charging time of electrons that are repelled from the particle \cite{Basha1989}. This slow charging time, coupled with the gradient in the equilibrium charge, is essentially why the particle oscillation amplitude can become so large in our experiments. 

Figure \ref{model_fit}a shows that the equilibrium charge gradient is $Q_\text{eq}'(0)/e\approx$ 20,000 electrons per millimeter. This is approximately 4-5 times larger than some reported measurements for similar particle sizes at higher pressures \cite{Douglass2012}. The discrepancy is likely due to the low pressures and plasma densities in our experiments. In Fig.\ \ref{efield_charge}c, the charge measurements are based on motion analysis of the particle, and a model for the electric field in the sheath. This is the largest unknown in our measurements. Without experimental characterization of spatial distribution of the sheath potential,  measurements of the particle charge can not be decoupled from the model of $E(z)$. For a given electric field profile, a smaller value of the charge gradient would necessarily require a higher value of $\nu$. Additionally, we note that since the particle mass $m$ can be combined with the charge $Q$ in the model, an uncertainty in the particle mass (Fig.\ \ref{PSD}) would not affect the fitted value of $\nu$, but it would affect the equilibrium charge profile (Fig.\ \ref{model_fit}a). In any case, delayed charging is the only proposed mechanism able to provide the necessary ``kick'' in the sheath to maintain oscillations with a consistent amplitude for minutes-long timescales. 


\section{Conclusion}

The spontaneous oscillation of micron-sized particles in a plasma sheath has been a topic of active research for more than 20 years. The oscillations represent a nonequilibrium mechanism where particles extract a net amount of energy from their environment, and can serve as the progenitors for emergent, many-body phenomena in dusty plasmas. Using a combination of high-speed video, Langmuir probe measurements, and numerical modeling, we investigated the largest-amplitude oscillations reported to date. The particle motion is strongly anharmonic, and the oscillation amplitude can reach more than 1 cm and is remarkably constant over minutes timescales. Several mechanisms have been invoked to explain the origin of spontaneous oscillations in different plasma environments, yet only delayed charging \cite{Nunomura1999,Ivlev2000Inf} is consistent with our observations.  The delayed charging model was able to reproduce the motion of the particles in our experiments with great accuracy, and provided a quantitative measurement of the characteristic charging frequency that is consistent with theoretical predictions in similar plasma conditions \cite{Basha1989}. We are not aware of any other methods to directly measure the charging frequency of particles in a plasma. 

It is important to reiterate the main source of error and future research directions from our work. The charge profile intrinsically depends on an accurate model for the electric field. Here we have used the Child-Langmuir law, yet the flatness of the charge profile deep into the sheath (Fig.\ \ref{efield_charge}c) suggest that the electron current to the particle must be larger than expected, possibly due to a higher electron temperature in the sheath \cite{Ingram1988}. We suspect that this enhancement in the sheath may be what produces the large gradient in charge that facilitates the large vertical oscillations. Another avenue of research that deserves further attention is the collective influence of multiple particles undergoing spontaneous oscillations since the local plasma environment is affected by the density of particles. 


\begin{acknowledgments} 
This work was supported by the NSF DMR Grant No. 1455086. 
\end{acknowledgments}

\end{document}